\documentclass[twocolumn,english,prb,showpacs]{revtex4-1}

\usepackage{float}
\usepackage[dvips,final]{graphicx}
\usepackage{color}
\usepackage{amsmath}

\usepackage{graphicx, subfigure}
\usepackage{pslatex}
\usepackage{relsize}
\usepackage{bm}
\usepackage{verbatim}

\usepackage{appendix}

\usepackage{booktabs}
\usepackage{xcolor}
\usepackage{soul}

\usepackage{natbib}

\begin{document}
\title{Spin-lattice model for cubic crystals}

\author{P. Nieves$^{1}$}
\email{Corresponding author: pablo.nieves.cordones@vsb.cz}
\author{J. Tranchida$^2$}
\author{S. Arapan$^{1}$}

\author{D. Legut$^{1}$}

\affiliation{$^1$ IT4Innovations, V\v{S}B - Technical University of Ostrava, 17. listopadu 2172/15, 70800 Ostrava-Poruba, Czech Republic}
\affiliation{$^2$ Computational Multiscale Department, Sandia National Laboratories, P.O. Box 5800, MS 1322, 87185 Albuquerque, NM, United States}

\date{\today}

\begin{abstract}

We present a methodology based on the N\'{e}el model to build a classical spin-lattice Hamiltonian for cubic crystals capable of describing  magnetic properties induced by the spin-orbit coupling like magnetocrystalline anisotropy and anisotropic magnetostriction, as well as exchange magnetostriction. Taking advantage of the analytical solutions of the N\'{e}el model, we derive theoretical expressions for the parameterization of the exchange integrals and N\'{e}el dipole and quadrupole terms that link them to the magnetic properties of the material. This approach allows to build accurate spin-lattice models with the desire magnetoelastic properties. We also explore a possible way to model the volume dependence of magnetic moment based on the Landau energy. This new feature can allow to consider the effects of hydrostatic pressure on the saturation magnetization. We apply this method to develop a spin-lattice model for BCC Fe and FCC Ni, and we show that it accurately reproduces the experimental elastic tensor,  magnetocrystalline anisotropy under pressure, anisotropic magnetostrictive coefficients, volume magnetostriction and saturation magnetization under pressure at zero-temperature. This work could constitute a step towards large-scale modeling of magnetoelastic phenomena.

\end{abstract}
\pagebreak
\maketitle

\section{Introduction}

Magnetoelastic interactions couple the motion of atoms in a magnetic material with atomic magnetic moments, and allow to transfer mechanical and thermal energies between phonon and magnon subsystems \cite{ma2020atomistic}. 
Magnetoelasticity is of great interest for applications, but also from a fundamental point of view. 
For instance, precise control of magnetization through a mechanical excitation of the motion of atoms in magnetic materials, and vice versa, has enabled the development of a wide range of technological applications such as sensors (torque sensors, motion and position sensors, force and stress sensors) and actuators (sonar transducer, linear motors, rotational motors, and hybrid magnetostrictive/piezoelectric devices)  \cite{sensors,sensors_actuators,actuator,Dapino}. 
Similarly, the combination of magnetism and heat is exploited in many applications like heat-assisted magnetic recording (HAMR) \cite{HAMR}, thermally assisted magnetic random access memory (MRAMs) \cite{TARAM}, ultrafast all-optically induced magnetization dynamics \cite{USD,Ostler2012}, magnetic refrigeration \cite{Kitanovski_2015}, and biomedical magnetic hyperthermia \cite{ABENOJAR2016440}.

Magnetoelastic effects can also have a strong influence on the thermo-mechanical properties of  materials. 
This is for example the case of the phononic component of the thermal conductivity.
Though magnon-phonon scattering, it can abruptly change through magnetic phase-transitions \cite{zhou2020atomistic,backlund1961experimental}.
For metallic oxides presenting strong magnetoelastic effects, and for which accurate thermal conductivity predictions can be of practical importance (such as uranium dioxide \cite{jaime2017piezomagnetism}), the development of accurate numerical models is still an ongoing process. Similarly, magnetoelastic effects can also play an important role in the thermal expansion of magnetic materials like in Invars, where  is large enough to cancel the normal
thermal contraction, leading to nearly zero
net thermal contraction over a broad
range of temperatures\cite{WASSERMAN1990237}.

Presently, the theoretical and modeling techniques have reached a great level of development and accuracy to describe the uncoupled dynamics of magnons and phonons at different spatial and time scales. Typically, in magnetic materials this is done by constraining or neglecting either the motion of
atomic magnetic moments or atoms. For example, in spin-polarized ab-inito molecular dynamics (AIMD) magnetic
moments are constrained in certain directions and only atomic positions are updated in each time step, while in
classical atomistic spin dynamics (SD) and molecular dynamics (MD) the motion of atoms or spins are neglected,
respectively \cite{Evans_2014,Eriksson_2017,PLIMPTON19951}. However, it is still a challenge to find suitable modeling approaches to deal with processes where
the interaction between magnons and phonons is essential, like in magneto-caloric and magneto-elastic phenomena.
The lack of such models is limiting the multi-scale design of materials suitable for relevant technological applications
based on these physical processes. Recently, novel attempts to address this problem have been proposed. 
Stockem \emph{et al.} demonstrated that for small supercells, a consistent interface can be designed to couple spin-polarized AIMD and classical SD \cite{Stockem}. 
Although offering an excellent level of accuracy, this approach presents the space and time scale limitations of first-principles approaches, and does not appear to be suited for running meso-scale magneto-elastic simulations.
Another concept, referred to as “spin-lattice dynamics”, is based on the combination of classical spin and molecular dynamics (SD-MD), which includes the spatial dependence of exchange integrals in the spin equation of motion, among other features \cite{Ma2008,Beaujouan2012,MA2016350,Wu2018,Fransson,Perera2017,TRANCHIDA2018406}.
The computational cost of this classical approach scales linearly with the number of magnetic atoms in the system \cite{TRANCHIDA2018406}. Combined to accurate massively parallel algorithms, this enables the simulation of multi-million magnetic atom systems on time scales sufficient to accurately study magnon-phonon relaxation processes \cite{TRANCHIDA2018406,MA2016350}.

These new ideas have opened up interesting opportunities and questions about how to model and study magneto-caloric and
magneto-elastic phenomena within a multi-scale approach.
In particular, the coarse-grained modeling of spin-orbit coupling (SOC) through magnetocrystalline anisotropy (MCA) in SD-MD
is currently a bottle neck of this issue \cite{Perera}. The single-ion model of MCA is widely used in SD, but unfortunately it does not couple atom and spin degrees of freedom. 
This drawback can be overcome using the N\'{e}el model (two-ion model) that reproduces the correct symmetry of MCA, and couples atom and spin motion. Hence, despite some limitations of the N\'{e}el model concerning non-magnetic atoms and its phenomenological nature \cite{Skomski}, it seems a promising starting point to build a SD-MD model capable of  simulating magneto-caloric and magneto-elastic phenomena. In this work, we propose a general procedure to find the parameters of the N\'{e}el model within the Bethe-Slater curve\cite{slater1,slater2,Sommerfeld_1933} that reproduces the MCA, isotropic and anisotropic magnetoelastic properties, and magnetization under pressure for cubic crystals at zero-temperature accurately.

\section{Methodology}
\label{section:method}

\subsection{Spin-Lattice Hamiltonian}
\label{section:SL}

In the following discussion, we consider the spin-lattice Hamiltonian 
\begin{equation}
\begin{aligned}
\mathcal{H}_{sl}(\boldsymbol{r},\boldsymbol{p},\boldsymbol{s}) & =  \mathcal{H}_{mag}(\boldsymbol{r},\boldsymbol{s})+\sum_{i=1}^N\frac{\boldsymbol{p}_i}{2m_i}+\sum_{i,j=1}^N\mathcal{V}(r_{ij}),
\label{eq:Ham_tot}
\end{aligned}
\end{equation}
where $\boldsymbol{r}_i$, $\boldsymbol{p}_i$, $\boldsymbol{s}_i$, and $m_i$ stand for the position, momentum, normalized magnetic moment and mass for each atom $i$ in the system, respectively, $\mathcal{V}(r_{ij})=\mathcal{V}(\vert \boldsymbol{r}_i-\boldsymbol{r}_j\vert)$ is the interatomic potential energy and $N$ is the total number of atoms in the system with total volume $V$. Here, we include the following interactions in the magnetic energy
\begin{equation}
\begin{aligned}
\mathcal{H}_{mag}(\boldsymbol{r},\boldsymbol{s}) & =  -\mu_{0}\sum_{i=1}^N\mu_i(v)\boldsymbol{H}\cdot\boldsymbol{s}_i -\frac{1}{2}\sum_{i,j=1,i\neq j}^N J(r_{ij})\boldsymbol{s}_i\cdot\boldsymbol{s}_j \\
& +\mathcal{H}_{L}(v) + \mathcal{H}_{N\acute{e}el}(\boldsymbol{r},\boldsymbol{s}),
\label{eq:Ham_mag0}
\end{aligned}
\end{equation}
where $\mu_{i}(v)$ is the atomic magnetic moment that depends on the volume per atom of the system $v=V/N$, $\mu_0$ is the vacuum permeability, $\boldsymbol{H}$ is the external magnetic field, $J(r_{ij})$ is the exchange parameter. The quantity $\mathcal{H}_{L}$ is the Landau energy\cite{Ma2012,Moruzzi,kubler,MA2016350}
\begin{equation}
\begin{aligned}
\mathcal{H}_{L}(v)=\sum_{i=1}^N(A_i \mu_i^2(v)+B_i \mu_i^4(v)+C_i \mu_i^6(v)),
\label{eq:landau_energy}     
\end{aligned}
\end{equation}
where $A_i$, $B_i$ and $C_i$ are parameters, while $\mathcal{H}_{N\acute{e}el}$ is the N\'{e}el interaction
\begin{equation}
\begin{aligned}
\mathcal{H}_{N\acute{e}el} & =   -\frac{1}{2}\sum_{i,j=1,i\neq j}^{N} \lbrace g(r_{ij}) + l_1(r_{ij})\left[ (\boldsymbol{e}_{ij}\cdot\boldsymbol{s}_i)(\boldsymbol{e}_{ij}\cdot\boldsymbol{s}_j)-\frac{\boldsymbol{s}_i\cdot\boldsymbol{s}_j}{3}\right]  \\
& + q_1(r_{ij})\left[ (\boldsymbol{e}_{ij}\cdot\boldsymbol{s}_i)^2-\frac{\boldsymbol{s}_i\cdot\boldsymbol{s}_j}{3}\right]\left[ (\boldsymbol{e}_{ij}\cdot\boldsymbol{s}_j)^2-\frac{\boldsymbol{s}_i\cdot\boldsymbol{s}_j}{3}\right]   \\
& + q_2(r_{ij})\left[ (\boldsymbol{e}_{ij}\cdot\boldsymbol{s}_i)(\boldsymbol{e}_{ij}\cdot\boldsymbol{s}_j)^3+(\boldsymbol{e}_{ij}\cdot\boldsymbol{s}_j)(\boldsymbol{e}_{ij}\cdot\boldsymbol{s}_i)^3\right]  \rbrace,
\label{eq:Neel_energy}
\end{aligned}
\end{equation}
where $\boldsymbol{e}_{ij}=\boldsymbol{r}_{ij}/r_{ij}$, and 
\begin{equation}
\begin{aligned}
l_1(r_{ij}) & = l(r_{ij})+\frac{12}{35}q(r_{ij}), \\
q_1(r_{ij}) & = \frac{9}{5}q(r_{ij}), \\
q_2(r_{ij}) & = -\frac{2}{5}q(r_{ij}). \\
\end{aligned}
\end{equation}
In the case of a collinear state ($\boldsymbol{s}_i\parallel\boldsymbol{s}_j$), the Eq. \ref{eq:Neel_energy} is reduced to
\begin{equation}
\begin{aligned}
\mathcal{H}_{N\acute{e}el} & =   - \frac{1}{2}\sum_{i,j=1,i\neq j}^{N} \lbrace g(r_{ij}) +l(r_{ij}) \left(\cos^2\psi_{ij} -\frac{1}{3}\right) \\
& + q(r_{ij}) \left( \cos^4\psi_{ij} -\frac{6}{7}\cos^2\psi_{ij} + \frac{3}{35}\right) \rbrace
\label{eq:Neel_energy_coll}
\end{aligned}
\end{equation}
where $\cos\psi_{ij}=\boldsymbol{e}_{ij}\cdot\boldsymbol{s}_i$.  The N\'{e}el energy reproduces the correct symmetry of MCA and magnetoelastic energy\cite{Skomski}. It is convenient to use $g(r_{ij})$ to offset the exchange and Landau energy in order to allow the forces and pressure become zero at the ground state,  as detailed in Ma \emph{et al.} \cite{Ma2008}. To do so, we write this function as
\begin{equation}
g(r_{ij})  = -J(r_{ij})+\frac{2}{N-1}(A_i \mu_i^2(v)+B_i \mu_i^4(v)+C_i \mu_i^6(v)).
\label{eq:g(r)0}
\end{equation}
This offset does not affect the precession dynamics of the spins. However, it allows to offset the corresponding mechanical forces. This particular choice of the offset also implies that the spatial dependence of the exchange and Landau energy is not taken into account in the evaluation of the magnetic energy at the ground state. The exchange and Landau energies determine the value of magnetic moments\cite{Ma2012}. In this model, we effectively take into account this fact by parameterizing the volume dependence of magnetic moment using first-principles calculations. The spatial dependence of the exchange and Landau energy would also contribute to the total energy when the lattice parameter is modified, influencing the energy versus volume curve from which the equation of state and elastic properties are derived. However, the lack of this contribution in the model should not compromise its accuracy as long as the interatomic potential $\mathcal{V}(r_{ij})$ correctly reproduces the equation of state and elastic properties. Typically, interatomic potentials are developed and designed for this purpose. As a result of this offset, we see that the second term in Eq.\ref{eq:g(r)0}  cancels with the Landau energy, so that we can simplify the magnetic Hamiltonian Eq.(\ref{eq:Ham_mag0}) by removing the Landau energy 
\begin{equation}
\begin{aligned}
\mathcal{H}_{mag}(\boldsymbol{r},\boldsymbol{s}) & =  -\mu_{0}\sum_{i=1}^N\mu_i(v)\boldsymbol{H}\cdot\boldsymbol{s}_i -\frac{1}{2}\sum_{i,j=1,i\neq j}^N J(r_{ij})\boldsymbol{s}_i\cdot\boldsymbol{s}_j \\
& + \mathcal{H}_{N\acute{e}el}(\boldsymbol{r},\boldsymbol{s}),
\label{eq:Ham_mag}
\end{aligned}
\end{equation}
and setting
\begin{equation}
g(r_{ij})  = -J(r_{ij}).
\label{eq:g(r)}
\end{equation}
Consequently, this approach has the advantage that it avoids the calculation of the parameters $A_i$, $B_i$ and $C_i$ in the Landau energy (Eq.\ref{eq:landau_energy}). As shown in Section \ref{section:mag_mom}, the parameterization of the volume dependence of magnetic moment might be simpler than the calculation of the parameters in the Landau energy\cite{Ma2012}. According to the N\'{e}el model, function $g(r_{ij})$ can be linked to the volume magnetostriction induced by the exchange interactions (isotropic magnetostriction)\cite{Chika}. The N\'{e}el dipole ($l(r_{ij})$) and quadrupole ($q(r_{ij})$) terms can describe the effects induced by SOC and crystal field interactions like  MCA and its strain dependence (anisotropic magnetostriction) \cite{Chika}. Here, we take into account the spatial dependence of $J(r_{ij})$, $l(r_{ij})$ and $q(r_{ij})$ using the Bethe-Slater curve, as implemented in the SPIN package of LAMMPS \cite{TRANCHIDA2018406}
\begin{equation}
\begin{aligned}
J(r_{ij}) & =  4\alpha_J \left(\frac{r_{ij}}{\delta_J}\right)^2 \left[1-\gamma_n\left(\frac{r_{ij}}{\delta_J}\right)^2\right] e^{-\left(\frac{r_{ij}}{\delta_J}\right)^2} \Theta(R_{c,J}-r_{ij}),\\
l(r_{ij}) & =  4\alpha_l \left(\frac{r_{ij}}{\delta_l}\right)^2 \left[1-\gamma_l\left(\frac{r_{ij}}{\delta_l}\right)^2\right] e^{-\left(\frac{r_{ij}}{\delta_l}\right)^2} \Theta(R_{c,l}-r_{ij}),\\
q(r_{ij}) & =  4\alpha_q \left(\frac{r_{ij}}{\delta_q}\right)^2 \left[1-\gamma_q\left(\frac{r_{ij}}{\delta_q}\right)^2\right] e^{-\left(\frac{r_{ij}}{\delta_q}\right)^2} \Theta(R_{c,q}-r_{ij}),
\label{eq:BS_J_g_q}
\end{aligned}
\end{equation}
where $\Theta(R_{c,n}-r_{ij})$ is the Heaviside step function and $R_{c,n}$ ($n=J,l,q$) is the cut-off radius. The parameters $\alpha_n$, $\gamma_n$, and $\delta_n$ ($n=J,l,q$) must be determined in order to reproduce the Curie temperature ($T_C$) and volume magnetostriction ($\omega_s$) via $J(r_{ij})$, as well as anisotropic magnetostriction and MCA through  $l(r_{ij})$ and $q(r_{ij})$. The parameterization of $J(r_{ij})$ with the Bethe-Slater curve is a well established procedure. For instance, to find the values of  $\alpha_J$, $\gamma_J$, and $\delta_J$, one can fit the Bethe-Slater curve to exchange parameters calculated with Density Functional Theory (DFT) at fixed equilibrium positions at zero-temperature \cite{TRANCHIDA2018406}. However, in some cases this procedure might lead to spin-lattice models that don't reproduce correctly either $T_C$ or $\omega_s$. Hence, a strategy to  parameterize $J(r_{ij})$ using the Bethe-Slater function in order to simulate correctly these properties is highly desirable. Similarly, the parameterization of $l(r_{ij})$ and $q(r_{ij})$ with the Bethe-Slater curve is a quite new approach, so that it is not clear how to obtain the values of these parameters yet. In Section \ref{section:BS}, we propose a general procedure to obtain these parameters for cubic crystals based on the theoretical analysis of the N\'{e}el model\cite{Chika}. In Section \ref{section:mag_mom} we explore a possible parameterization of the volume dependence of magnetic moment using the Landau energy.  In the present work, we study this model only at zero-temperature. The equations of motion of this model at finite-temperature are those implemented in the SPIN package of LAMMPS\cite{TRANCHIDA2018406,PLIMPTON19951}. A detailed description of these equations can be found in Ref. \onlinecite{TRANCHIDA2018406}.

\subsection{Procedure to calculate the Bethe-Slater parameters of N\'{e}el interaction for cubic crystals}
\label{section:BS}

The basic idea to calculate the Bethe-Slater parameters for the N\'{e}el interaction is to find the theoretical relations that link Eq. (\ref{eq:Neel_energy_coll}) to both the MCA and magnetoelastic energies. To illustrate this method, we will apply it to simple cubic (SC), body-centered cubic (BCC) and face-centered cubic (FCC) crystals. The MCA energy for cubic systems reads \cite{Handley}
\begin{equation}
\begin{aligned}
    \mathcal{H}_{MCA}^{cub} (\boldsymbol{\alpha},r)= VK_1(r)(\alpha_x^2\alpha_y^2+\alpha_x^2\alpha_z^2+\alpha_y^2\alpha_z^2), 
\label{eq:E_mca_cub}     
\end{aligned}
\end{equation}
where $K_1$ is the first MCA constant with units of energy per volume, $r$ is the distance to the first nearest neighbor, $V$ is the volume of the system, and $\alpha_i$ ($i=x,y,z$) are the direction cosines of magnetization. From this equation we have
\begin{equation}
\begin{aligned}
    VK_1 (r) = 4\left[\mathcal{H}_{MCA}^{cub} \left(\frac{1}{\sqrt{2}},\frac{1}{\sqrt{2}},0,r\right)-\mathcal{H}_{MCA}^{cub} (1,0,0,r)\right]. 
\label{eq:E_mca_cub_K}     
\end{aligned}
\end{equation}
Next,  we evaluate the Eq. \ref{eq:Neel_energy_coll} with magnetic moment directions $\boldsymbol{s}=\left(\frac{1}{\sqrt{2}},\frac{1}{\sqrt{2}},0\right)$ and $\boldsymbol{s}=\left(1,0,0\right)$ up to first nearest neighbors, and we replace it in Eq. \ref{eq:E_mca_cub_K} in order to ensure that the N\'{e}el energy gives the correct MCA energy. And by doing so,  we find the following relations for SC, BCC and FCC \cite{Chuang}
\begin{equation}
\begin{aligned}
& SC: q(r_0) = \frac{V_0K_1(r_0)}{2N}=\frac{1}{2}r_0^3K_1(r_0),\\
& BCC: q(r_0) = -\frac{9 V_0K_1(r_0)}{16N}=-\frac{\sqrt{3}}{4}r_0^3K_1(r_0),\\
& FCC: q(r_0) = -\frac{ V_0K_1(r_0)}{N}=-\frac{1}{\sqrt{2}}r_0^3K_1(r_0),
\label{eq:E_mca_cub_K_q}     
\end{aligned}
\end{equation}
where $r_0$ is the equilibrium distance to the first nearest neighbors, and $N$ is the number of atoms in the equilibrium volume $V_0$. Here, $q(r_0)$ has units of energy per atom. In Appendix \ref{App_dqdr} we show that the derivative of $q(r)$ with respect to $r$ can be written as
\begin{equation}
\begin{aligned}
& SC: r_0\frac{\partial q}{\partial r}\Big\vert_{r=r_0}  = \frac{3}{2}r_0^3K_1(r_0)\left[1-\frac{B}{K_1}\frac{\partial K_1}{\partial P}\right]_{r=r_0},\\
& BCC: r_0\frac{\partial q}{\partial r}\Big\vert_{r=r_0}  = - \frac{3\sqrt{3}}{4}r_0^3K_1(r_0)\left[1-\frac{B}{K_1}\frac{\partial K_1}{\partial P}\right]_{r=r_0},\\
& FCC: r_0\frac{\partial q}{\partial r}\Big\vert_{r=r_0}  = - \frac{3}{\sqrt{2}}r_0^3K_1(r_0)\left[1-\frac{B}{K_1}\frac{\partial K_1}{\partial P}\right]_{r=r_0},
\label{eq:E_mca_cub_K_dqdr}     
\end{aligned}
\end{equation}
where $B$ is the bulk modulus and $P$ is pressure. Here again, $r_0\partial q/\partial r$ has units of energy per atom. Note that the dipole term in Eq. \ref{eq:Neel_energy_coll} cancels out after summing all first nearest neighbors, so that it does not contribute to the MCA in the cubic crystal symmetry. Since we are only considering N\'{e}el interactions up to the first nearest neighbors, we set the cut-off radius $R_{c,q}$ in between the first and second nearest neighbors in Eq. \ref{eq:BS_J_g_q}, that is
\begin{equation}
\begin{aligned}
q(r_0)  =  4\alpha_q \left(\frac{r_0}{\delta_q}\right)^2 \left[1-\gamma_q\left(\frac{r_0}{\delta_q}\right)^2\right] e^{-\left(\frac{r_0}{\delta_q}\right)^2}.
\label{eq:BS_q}
\end{aligned}
\end{equation}
The derivative of this function with respect to $r$ is
\begin{equation}
\begin{aligned}
 \frac{\partial q}{\partial r}\Big\vert_{r=r_0} = \frac{8\alpha_qr_0 e^{-\left(\frac{r_0}{\delta_q}\right)^2}}{\delta_q^6}\left[\gamma_qr_0^4-(1+2\gamma_q)\delta_q^2r_0^2 +\delta_q^4 \right].
\label{eq:d_BS}     
\end{aligned}
\end{equation}
Hence, we have two equations with three unknown variables $\alpha_q$, $\gamma_q$, and $\delta_q$. A reasonable strategy to reduce the number of unknown variables is to set $\delta_q$ equal to the equilibrium distance to the first nearest neighbors $r_0$ ($\delta_q=r_0$) because it has unit of distance and can be easily estimated. Hence, solving Eqs. \ref{eq:BS_q} and \ref{eq:d_BS}  gives 
\begin{equation}
\begin{aligned}
\delta_q & = r_0, \\
\alpha_q &  = \frac{ e}{8}\left[2q(r_0) - r_0\frac{\partial q}{\partial r}\Big\vert_{r=r_0} \right], \\
\gamma_q &  = \frac{ r_0\frac{\partial q}{\partial r}\Big\vert_{r=r_0}}{r_0\frac{\partial q}{\partial r}\Big\vert_{r=r_0} - 2q(r_0)  }.
\label{eq:BS_q_dq}     
\end{aligned}
\end{equation}
These are the Bethe-Slater parameters in terms of $K_1$ and $\partial K_1/\partial P$ (via Eqs. \ref{eq:E_mca_cub_K_q} and \ref{eq:E_mca_cub_K_dqdr}) to model the physics of MCA within the N\'{e}el model. 

Let's now find the values of the Bethe-Slater parameters that simulate magnetostriction. The magnetoelastic energy for cubic systems (point groups $432$, $\bar{4}3m$, $m\bar{3}m$) reads  \cite{CLARK1980531,Cullen}
\begin{equation}
\begin{aligned}
    \frac{\mathcal{H}_{me}^{cub}}{V_0}  & =  b_0(\epsilon_{xx}+\epsilon_{yy}+\epsilon_{zz})+b_1(\alpha_x^2\epsilon_{xx}+\alpha_y^2\epsilon_{yy}+\alpha_z^2\epsilon_{zz})\\
    & +  2b_2(\alpha_x\alpha_y\epsilon_{xy}+\alpha_x\alpha_z\epsilon_{xz}+\alpha_y\alpha_z\epsilon_{yz}), 
\label{eq:E_me_cub_I}     
\end{aligned}
\end{equation}
where $b_0$, $b_1$ and $b_2$ are the magnetoelastic constants with units of energy per volume, and $\epsilon_{ij}$ are the elements of the strain tensor. For small deformations (infinitesimal strain theory), the strain tensor can be expressed in terms of the displacement vector $\boldsymbol{u}$ as\cite{Landau,Maelas}
\begin{equation}
\begin{aligned}
     \epsilon_{ij}=\frac{1}{2}\left(\frac{\partial u_{i}}{\partial r_{j}}+\frac{\partial u_j}{\partial r_i}\right),\quad\quad i,j=x,y,z
    \label{eq:disp_vec}
\end{aligned}
\end{equation}
where $\partial u_i/ \partial r_j$ is called the displacement gradient. For this definition of the strain tensor, the elastic energy for cubic crystal reads\cite{Landau,Maelas}
\begin{equation}
\begin{aligned}
\frac{\mathcal{H}_{el}^{cub}}{V_0} & = \frac{c_{11}}{2}(\epsilon_{xx}^2+\epsilon_{yy}^2+\epsilon_{zz}^2)+c_{12}(\epsilon_{xx}\epsilon_{yy}+\epsilon_{xx}\epsilon_{zz}+\epsilon_{yy}\epsilon_{zz})\\
& + 2c_{44}(\epsilon_{xy}^2+\epsilon_{yz}^2+\epsilon_{xz}^2),
\end{aligned}
\label{eq:E_el_cub}
\end{equation}
where $c_{11}$, $c_{12}$ and $c_{44}$ are the elastic constants. After evaluating the N\'{e}el energy (Eq.\ref{eq:Neel_energy_coll}) for a strained cubic crystal up to first nearest neighbors, and equalizing it to Eq.\ref{eq:E_me_cub_I}, one finds for SC, BCC and FCC \cite{Chika,Handley}
\begin{equation}
\begin{aligned}
& SC:  \; l(r_0) = -\frac{V_0 b_2}{2 N},\;\;\; r_0\frac{\partial l}{\partial r}\Big\vert_{r=r_0} =  -\frac{V_0 b_1}{ N} ,\\
& BCC: \;  l(r_0) =  -\frac{3V_0 b_1}{8 N},\;\;\; r_0\frac{\partial l}{\partial r}\Big\vert_{r=r_0} = \frac{3V_0 }{8 N}(b_1-3b_2) ,\\
& FCC:\;  l(r_0) = \frac{V_0}{2 N}\left(\frac{b_2}{2}-b_1\right),\; r_0\frac{\partial l}{\partial r}\Big\vert_{r=r_0} = \frac{V_0}{ N}\left(b_1-\frac{3b_2}{2}\right).
\label{eq:l_dl}     
\end{aligned}
\end{equation}
Here, we neglected the quadrupole contribution to the magnetoelastic energy \cite{Chuang}. This approximation is reasonable when $q(r_0)\ll l(r_0)$. In Section \ref{section:SL_Fe_Ni}, we show that BCC Fe and FCC Ni fulfill this condition.  Next, as we did previously, inserting Eq.\ref{eq:l_dl} into the Bethe-Slater curve and its derivative, and setting $\delta_l=r_0$ allow us to obtain
\begin{equation}
\begin{aligned}
\delta_l & = r_0, \\
\alpha_l &  = \frac{ e}{8}\left[2l(r_0) - r_0\frac{\partial l}{\partial r}\Big\vert_{r=r_0} \right], \\
\gamma_l &  = \frac{ r_0\frac{\partial l}{\partial r}\Big\vert_{r=r_0}}{r_0\frac{\partial l}{\partial r}\Big\vert_{r=r_0} - 2l(r_0)  }.
\label{eq:BS_l_dl}     
\end{aligned}
\end{equation}
These are the Bethe-Slater parameters in terms of $b_1$ and $b_2$ (via Eq. \ref{eq:l_dl}) to model the anisotropic magnetostriction within the N\'{e}el model. 

Lastly, we show the parameterization of the exchange interaction via the first term in the N\'{e}el model (Eq.\ref{eq:g(r)}) to simulate $T_{C}$ and $\omega_s$. From the analysis of the N\'{e}el model up to first nearest-neighbours one finds\cite{Chika}
\begin{equation}
\begin{aligned}
& SC:  \; J(r_0) = \frac{k_BT_C}{2},\;\;\; r_0\frac{\partial J}{\partial r}\Big\vert_{r=r_0} =  \frac{\omega_s (c_{11}+2c_{12}) V_0}{ 3N} ,\\
& BCC: \;  J(r_0) =  \frac{3k_BT_C}{8},\;\;\; r_0\frac{\partial J}{\partial r}\Big\vert_{r=r_0} = \frac{\omega_s (c_{11}+2c_{12}) V_0}{ 4N},\\
& FCC:\;  J(r_0) = \frac{k_BT_C}{4},\; r_0\frac{\partial J}{\partial r}\Big\vert_{r=r_0} = \frac{\omega_s (c_{11}+2c_{12}) V_0}{ 6N},
\label{eq:J_dJ}     
\end{aligned}
\end{equation}
where $k_B$ is the Boltzmann constant. The relation between $J(r_0)$ and $T_C$ was obtained using the Mean-Field Approximation (MFA). Inserting Eq.\ref{eq:J_dJ} into the Bethe-Slater curve and its derivative, and setting $\delta_J=r_0$ allow us to obtain
\begin{equation}
\begin{aligned}
\delta_J & = r_0, \\
\alpha_J &  = \frac{ e}{8}\left[2J(r_0) - r_0\frac{\partial J}{\partial r}\Big\vert_{r=r_0} \right], \\
\gamma_J &  = \frac{ r_0\frac{\partial J}{\partial r}\Big\vert_{r=r_0}}{r_0\frac{\partial J}{\partial r}\Big\vert_{r=r_0} - 2J(r_0)  },
\label{eq:BS_J_dJ}     
\end{aligned}
\end{equation}
where the cut-off radius $R_{c,J}$ must be in between the first and second nearest neighbors. Notice that a similar procedure to this one could be used if a function with three parameters different to the Bethe-Slater curve is chosen to describe the functions $l(r)$, $q(r)$ and $J(r)$ in SD-MD simulations.

\subsection{Volume dependence of magnetic moment}
\label{App_magmom}

A widely used approximation in SD consists to constrain the magnitude of atomic magnetic moments to a constant value \cite{Evans_2014,Eriksson_2017}, which is a good approximation for many practical applications. However, the magnitude of the atomic magnetic moments can change significantly when the volume of the system changes greatly. This effect can be analyzed in terms of the Landau expansion  around the critical point where the magnetization becomes zero \cite{LANDAU1984130}. The Landau expansion contains only even powers of magnetization to fulfill the time-reversal symmetry. For instance, the Curie temperature ($T_c$) is a critical point where the magnetization becomes zero due to the disorder of magnetic moments induced by thermal fluctuations. Another critical point is the volume per atom $v_c$ where magnetic moment collapses ($\mu(v_c)=0$)\cite{Moruzzi}. Moruzzi identified three types of transitions for the magnetic moment collapse\cite{Moruzzi,kubler}. Here, we discuss about type I transition where the behavior is continuous across $v_c$. In particular, we explore a possible parameterization of the magnetic moments based on the Landau energy to take into account its volume dependence. For a system with $N$ atoms with equal magnetic moments $\mu$ we can write the Landau expansion close to $v_c$ as
\begin{equation}
\begin{aligned}
\mathcal{H}_{L}(v) & =\sum_{i=1}^N(A_i \mu_i^2(v)+B_i\mu_i^4(v))+O( \mu^6)\\
& = N (A\mu^2(v)+B\mu^4(v)),
\label{eq:landau_energy_app}     
\end{aligned}
\end{equation}
where $A$ and $B$ are parameters, and $v$ is the volume per atom of the system. The analysis of the Landau expansion yields\cite{Moruzzi,kubler}
\begin{equation}
\begin{aligned}
\mu(v) \propto \sqrt{v-v_c}.
\label{eq:magmom_1}     
\end{aligned}
\end{equation}
This square root dependence describes well the magnetic moment behaviour very close to $v_c$. However, for many practical applications the equilibrium volume is significantly far from $v_c$, so that one needs to include additional terms in Eq.\ref{eq:magmom_1}. In order to do so, we perform a Taylor expansion of the square of magnetic moment $\mu^2$ around $v_c$, that is
\begin{equation}
\begin{aligned}
\mu^2(v) & = \mu^2(v_c)+\frac{\partial \mu^2}{\partial v}\Bigg\vert_{v=v_c}(v-v_c) \\
& +\frac{1}{2}\frac{\partial^2 \mu^2}{\partial v^2}\Bigg\vert_{v=v_c}(v-v_c)^2 \\
& +\frac{1}{6}\frac{\partial^3 \mu^2}{\partial v^3}\Bigg\vert_{v=v_c}(v-v_c)^3+O((v-v_c)^4),
\label{eq:magmom_taylor}     
\end{aligned}
\end{equation}
where $\mu^2(v_c)=0$. Making a square root on both sides of this equation we have
\begin{equation}
\begin{aligned}
\mu(v) & = \sqrt{\alpha_{\mu}(v-v_{c})+\beta_{\mu}(v-v_{c})^2+\gamma_{\mu}(v-v_{c})^3}\\
& \cdot \Theta(v-v_{c}),
\label{eq:magmom_final}     
\end{aligned}
\end{equation}
where
\begin{equation}
\begin{aligned}
\alpha_{\mu} & = \frac{\partial \mu^2}{\partial v}\Bigg\vert_{v=v_c} \\
\beta_{\mu} & = \frac{1}{2}\frac{\partial^2 \mu^2}{\partial v^2}\Bigg\vert_{v=v_c} \\
\gamma_{\mu} & = \frac{1}{6}\frac{\partial^3 \mu^2}{\partial v^3}\Bigg\vert_{v=v_c}.
\label{eq:magmom_parameter}     
\end{aligned}
\end{equation}
The Heaviside step function $\Theta(v-v_{c})$ was introduced in Eq. \ref{eq:magmom_final} to ensure that the magnetic moment is zero at volumes lower than $v_c$. The Taylor expansion was considered up to the third order which is enough to correctly describe  the magnetic moment of BCC Fe and FCC Ni within the range of volume per atom discussed in this work ($v<20$\r{A}$^3$/atom). For cases with larger volume per atom than $20$\r{A}$^3$/atom one might need to include higher order terms in the Taylor expansion. In the vicinity of the critical volume ($(v-v_c)/v_c\ll 1$) Eq. \ref{eq:magmom_final} becomes Eq.\ref{eq:magmom_1}, so that the result derived from the Landau expansion is recovered\cite{Moruzzi}. 

\begin{figure}[h]
\centering
\includegraphics[width=\columnwidth ,angle=0]{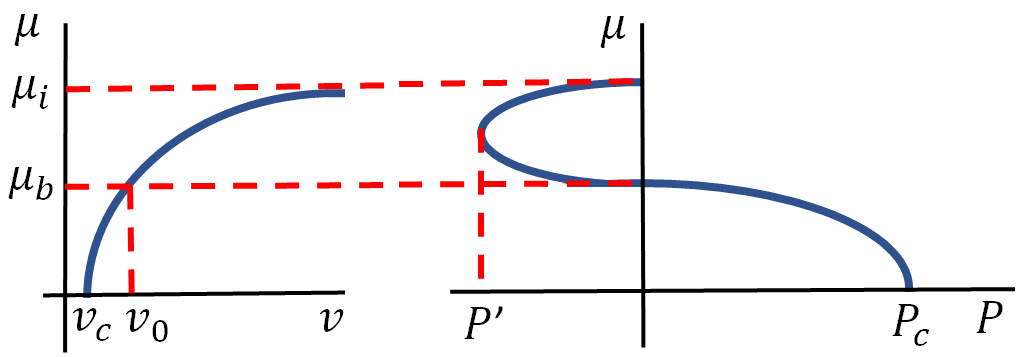}
\caption{Schematic of the volume and hydrostatic pressure dependence of magnetic moment. Symbols $\mu_b$ and $\mu_i$ represent the magnetic moment for bulk at zero-pressure and isolated atom, respectively. }
\label{fig:diagram_m_v_p}
\end{figure}

Alternatively, instead of considering the volume dependence for the parameterization of the magnetic moment $\mu(v)$, one could consider the pressure dependence of magnetic moment $\mu(P)$. In this case, one could apply the same procedure but now performing the Taylor expansion around the critical pressure $P_c$ where the magnetic moment collapses. Note that the function $\mu(P)$ could only be evaluated in this way from $P_c$ up to the negative pressure $P'$ at which the pressure is reversed due to the large interatomic distance, see Fig. \ref{fig:diagram_m_v_p}.  In this model, longitudinal fluctuations of magnetic moments at finite temperature would naturally emerge from the fluctuations of the volume per atom or pressure. 

\section{Spin-lattice model for BCC Fe and FCC Ni}
\label{section:SL_Fe_Ni}

In this section, we build a SD-MD model for BCC Fe and FCC Ni based on the methodology presented in Section \ref{section:method}. The construction of the model is splitted into the following stages in order to systematically compute each term in Eq.\ref{eq:Ham_tot}, where the magnetic Hamiltonian is given by Eq.\ref{eq:Ham_mag}.

\subsection{Interatomic potential}
\label{section:force_field}

\begin{figure}[h]
\centering
\includegraphics[width=\columnwidth ,angle=0]{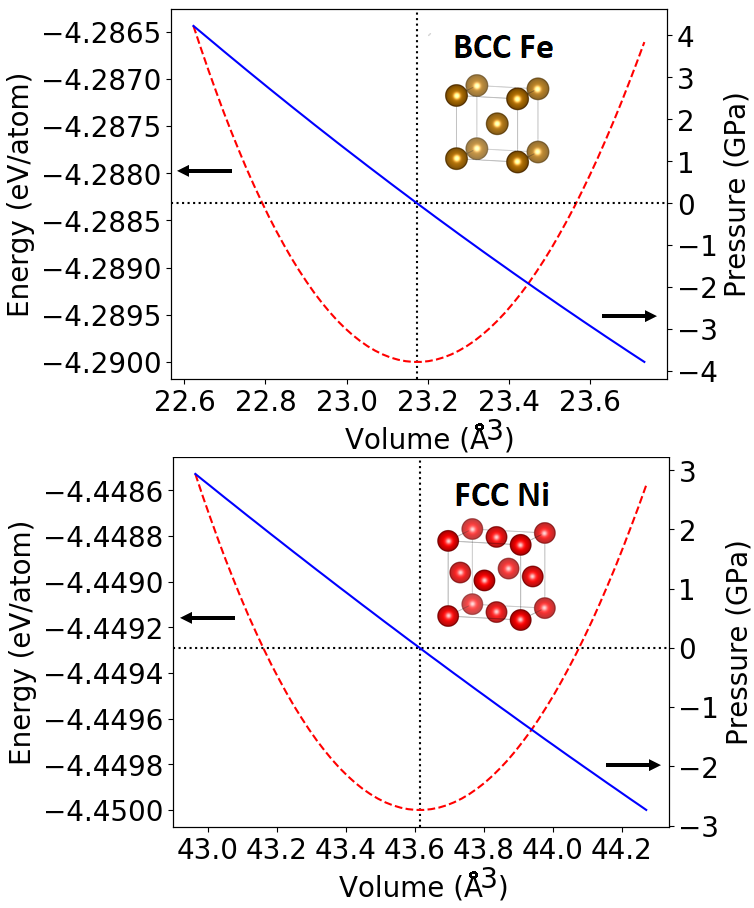}
\caption{Calculation of the equation of the state for (top) BCC Fe and (bottom) FCC Ni with the SD-MD model including only the MEAM potential.}
\label{fig:eos}
\end{figure}

 In the model we set the modified embedded atom method (MEAM) potentials developed by Asadi et al.\cite{asadi2015} and Lee et al.\cite{Lee2003} for the interatomic potential $\mathcal{V}(r_{ij})$ of BCC Fe and FCC Ni, respectively. These potentials give an elastic tensor very close to the experimental one at zero-temperature. In this first stage, it is convenient to find the equilibrium volume and bulk modulus given by the model including only the MEAM potential. To do so, we compute the energy of a set of conventional unit cells with different volume using the software LAMMPS\cite{PLIMPTON19951} with the SPIN package\cite{TRANCHIDA2018406}, and we fit it to the Murnaghan equation of state (EOS) \cite{Murnaghan1944,Fu}. We verify that the pressure in the selected equilibrium state is lower than $5\times10^{-5}$ GPa.  In Fig.\ref{fig:eos}, we present the calculation of the energy versus volume curve for the conventional unit cell (2 atoms/cell for BCC Fe and 4 atoms/cell for FCC Ni). The equilibrium volume and bulk modulus found with this procedure is  $v_0=11.586754$ \r{A}$^3$/atom and $B=166.73$ GPa for BCC Fe, and   $v_0=10.903545$ \r{A}$^3$/atom and $B=188.85$ GPa for FCC Ni. Hence, the equilibrium distance to the first nearest neighbor is $r_0=2.4690386$ \r{A} for BCC Fe, and $2.4890153$ \r{A} for FCC Ni. At this stage, it is also convenient to compute the elastic constants. To do so, we evaluate the elastic tensor with software AELAS\cite{AELAS} interfaced with LAMMPS at the equilibrium volume $v_0$ including the MEAM potential. The developed interface between AELAS and LAMMPS is available on GitHub repository \cite{aelas_lmp}. Here, we make use of the program Atomsk to convert some input files \cite{atomsk}. The calculated values and experimental ones are shown in Table \ref{table:data_properties}. We see that these interatomic potentials gives a very similar elastic tensor to the experiment.

\subsection{Magnetic moment}
\label{section:mag_mom}

Next, we find the parameterization of the volume dependence of magnetic moment $\mu(v)$. Here, we estimate it using DFT. Namely, the parameters $v_c$, $\alpha_\mu$, $\beta_\mu$ and $\gamma_\mu$ in Eq.\ref{eq:magmom_final} are obtained by fitting this equation to the magnetic moment versus volume curve given by DFT. The DFT calculations are performed with VASP code \cite{vasp_1,vasp_2,vasp_3}, which is an implementation of the projector augmented wave (PAW) method \cite{vasp_4}. We use the interaction potentials generated for the Perdew-Burke-Ernzerhof (PBE) version \cite{Perdew} of the Generalized Gradient Approximation (GGA).   We set an automatic Monkhorst-Pack k-mesh \cite{Monk} gamma-centered grid with length parameter $R_k=60$. The interactions were described by a PAW potential with 14 and 16 valence electrons for BCC Fe and FCC Ni, respectively. 
\begin{figure}[h]
\centering
\includegraphics[width=\columnwidth ,angle=0]{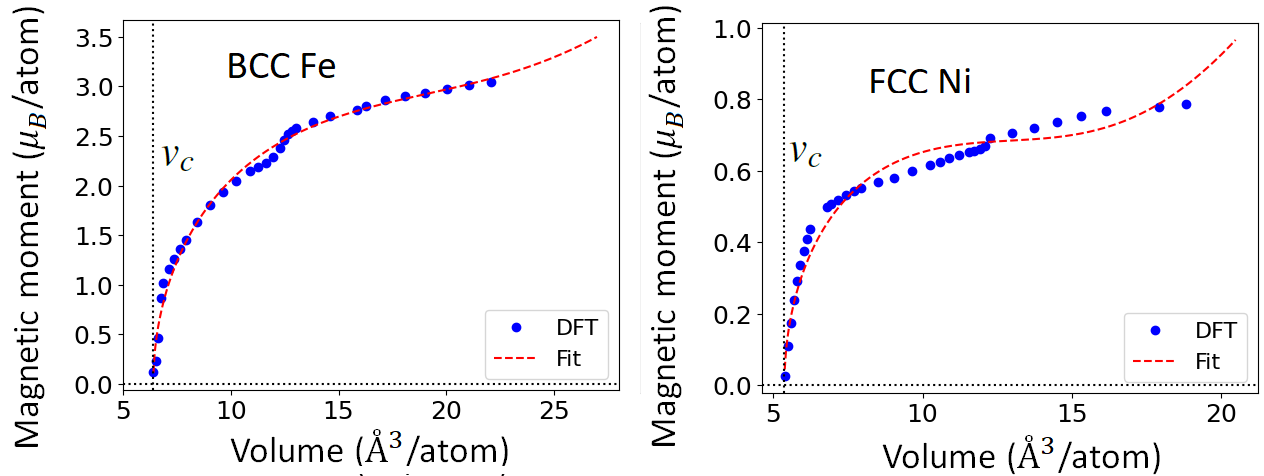}
\caption{Calculation of the magnetic moment versus volume under normal deformations obtained with DFT (blue dots) for BCC Fe and FCC Ni. Red line stands for the fitting curve.}
\label{fig:landau}
\end{figure}

The results of these calculations and corresponding fitting curves are shown in Fig.\ref{fig:landau}. Very similar results were previously reported by Moruzzi et al. using the augmented spherical wave (ASW) method\cite{MORUZZI199397}. We see that the form of Eq.\ref{eq:magmom_final}  describes quite well the data obtained by DFT. In the case of FCC Ni the deviation between the fitted curves and DFT data is slightly larger than for BCC Fe. A better fit could be achieved by adding higher order terms in Eq.\ref{eq:magmom_taylor}. The values of the fitting parameters $v_c$, $\alpha_\mu$, $\beta_\mu$ and $\gamma_\mu$ are presented in Table \ref{table:data_BS}.  Inserting these values into Eq.\ref{eq:magmom_final} allows us to compute the magnetic moment at the equilibrium volume given by the SD-MD model including only MAEM potential obtained in Section \ref{section:force_field} ($v_0=11.586754$ \r{A}$^3$/atom for BCC Fe, and $v_0=10.903545$ \r{A}$^3$/atom for FCC Ni). This calculation gives $\mu(v_0)=2.34\mu_B$ for Fe, and $\mu(v_0)=0.67\mu_B$ for Ni, while the experimental values are $2.22\mu_B$ and $0.606\mu_B$ for Fe and Ni, respectively \cite{Handley}. We see that this procedure overestimates slightly the magnetic moment.

\begin{figure}[h]
\centering
\includegraphics[width=\columnwidth ,angle=0]{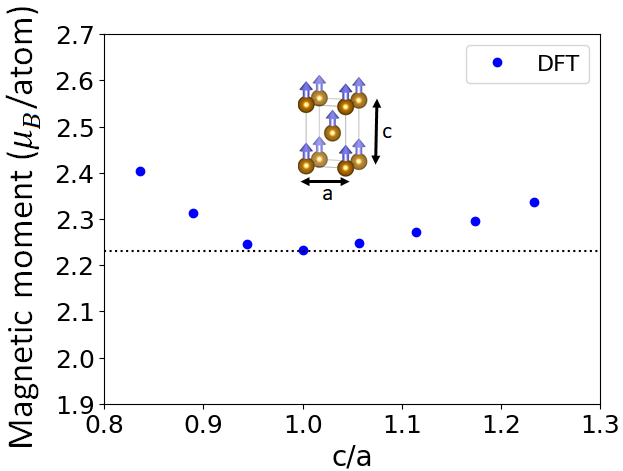}
\caption{Magnetic moment under a volume-conserving tetragonal deformation of BCC Fe ($c/a=1$) calculated with DFT.}
\label{fig:c_a_mag}
\end{figure}

The volume dependence of magnetic moment will allow us to study how magnetization changes under hydrostatic pressure (normal deformation). Note that in this model the magnitude of the magnetic moment will not change under volume-conserving deformations. To check the validity of this approximation, we run some additional DFT calculations with VASP using the same setting as before to obtain the magnetic moment under volume-conserving tetragonal deformation\cite{nieves2020maelas} for BCC Fe. The results are plotted in Fig.\ref{fig:c_a_mag}. We observe that a significant change of the magnetic moment only takes place at large tetragonal deformations. We verify that a similar trend is also observed for other types of volume-conserving deformations like trigonal deformation\cite{nieves2020maelas}. Therefore, to some extent, the model might be able to describe the behaviour of magnetic moment and magnetization under  deformations that combines an arbitrarily large  normal deformation (which changes the volume preserving the cubic symmetry) with a small volume-conserving deformation that changes the crystal symmetry.

\subsection{Exchange interaction}
\label{section:exchange}

Let's now compute the parameterization of $J(r)$. Firstly, note that the equilibrium interatomic distance, EOS and elastic constants of the ground state (collinear state)  are unchanged after the exchange interaction is added to the SD-MD model thanks to the offset in the exchange energy. It is interesting to analyze the influence of  different types of parameterization of $J(r)$ on the volume magnetostriction $\omega_s$. Hence, for the parameterization of $J(r)$ we consider the following two sets of parameters for $\alpha_J$, $\gamma_J$, $\delta_J$ and $R_{c,J}$.

\subsubsection{Set I: Effective short range exchange}

The set I is calculated following the procedure described in Section \ref{section:BS}, so that it leads to an effective short range exchange interaction.
As mentioned above, the equilibrium distance to the first nearest neighbor at the ground state is not changed by the exchange interaction due to the offset in the exchange energy, so that according to Eq.\ref{eq:BS_J_dJ} we set $\delta_J^{(I)}=r_0=2.4690386$ \r{A} for BCC Fe, and $\delta_J^{(I)}=2.4890153$ \r{A} for FCC Ni. Next, we see in Eq. \ref{eq:J_dJ} that we need as inputs $T_C$, $c_{11}$, $c_{12}$ and $\omega_s$ to compute the $J(r_0)$ and $\partial J/\partial r$. In general, these inputs can be obtained by theory or experiment. For instance, here we use the experimental value of $T_C$ (1043 K for BCC Fe and 627 K for FCC Ni)\cite{Handley}. For the elastic constants, we will make use of the theoretical values obtained by the SD-MD model itself using only the MEAM potential (see Table \ref{table:data_properties}). The experimental measurement of volume magnetostriction is difficult, and one can find significant discrepancies between different works\cite{shimizu1978,WASSERMAN1990237}. Hence, we will use the theoretical value of $\omega_s$ at zero-temperature calculated by Shuimizu using the itinerant electron model\cite{shimizu1978}, that is $\omega_s=1.16\times10^{-2}$ for BCC Fe and $3.75\times10^{-4}$ for FCC Ni. Inserting these quantities into Eq.\ref{eq:BS_J_dJ} via Eq.\ref{eq:J_dJ} gives $\alpha_J^{(I)}=-12.5921$ meV/atom and $\gamma_J^{(I)}=2.81897$ for BCC Fe, and $\alpha_J^{(I)}=8.35847$ meV/atom and $\gamma_J^{(I)}=-0.098217$ for FCC Ni.

\subsubsection{Set II: Long range exchange}

\begin{figure}[h]
\centering
\includegraphics[width=\columnwidth ,angle=0]{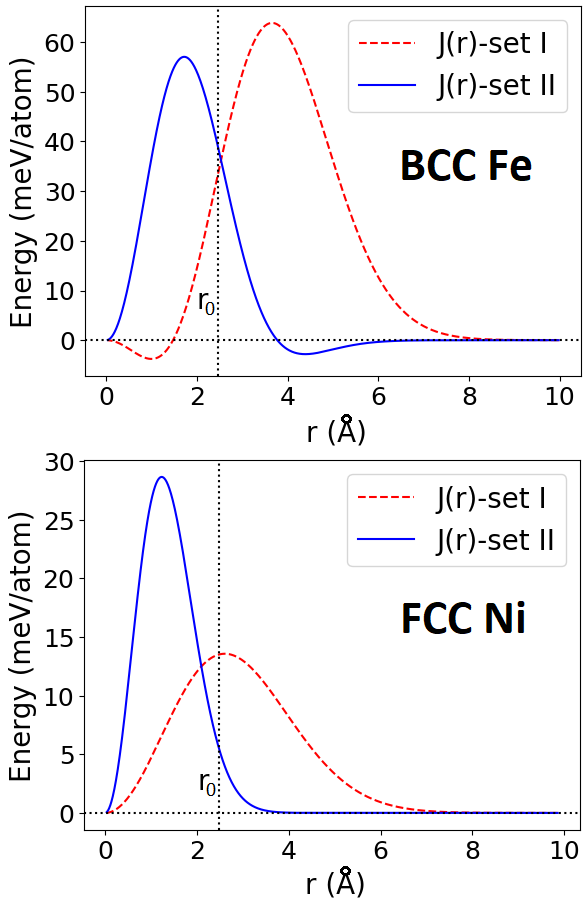}
\caption{Calculation of the Bethe-Slater function $J(r)$ and $q(r)$ for (top) BCC Fe and (bottom) FCC Ni using the two set of parameters given in Table \ref{table:data_BS}. Vertical dash line stands for the equilibrium distance of the first nearest neighbors $r_0$.}
\label{fig:BS_J_Fe_Ni}
\end{figure}

The second set of parameters (set II) is obtained by fitting the Bethe-Slater function to the exchange integrals given by first-principles calculations\cite{TRANCHIDA2018406}. The fitted parameters ($\alpha_J^{(II)}$, $\gamma_J^{(II)}$ and $\delta_J^{(II)}$) are shown in Table \ref{table:data_BS}. The value of $\alpha_J^{(II)}$ taken from Ref.\cite{TRANCHIDA2018406} has been multiplied by $2$ due to the factor $1/2$ in the exchange energy given by Eq.\ref{eq:Ham_mag}. Here, we set a large cut-off $R_{c,J}^{(II)}=4.5$ \r{A} to take into account the exchange interactions beyond first nearest neighbors (long range exchange interaction). The Bethe-Slater function with parameters from set I and II is plotted in Fig.\ref{fig:BS_J_Fe_Ni}. This figure will be analyzed in the context of volume magnetostriction in Section \ref{section:vol_mag}.

\subsection{N\'{e}el energy}
\label{section:neel_energy}

Now, we are in a position to calculate the Bethe-Slater parameters for the dipole and quadrupole terms of the N\'{e}el interaction given by Eqs.\ref{eq:BS_l_dl} and \ref{eq:BS_q_dq}, respectively. Firstly, we notice that a key quantity in these equations is the equilibrium distance to the first nearest neighbors $r_0$, which obviously depends on the N\'{e}el interaction. Fortunately, the energy of the dipole and quadrupole terms of the N\'{e}el interaction for Fe and Ni are of the order of $\mu$eV/atom (see Fig.\ref{fig:neel_test}), so that they are much lower than the total energy (eV/atom). As a result, these terms only induce a very small change in $r_0$ when they are included in the SD-MD model. This fact allows us to use $r_0$ given by the SD-MD model including only the MEAM potential and exchange interaction to calculate the Bethe-Slater parameters for the dipole and quadrupole terms of the N\'{e}el interaction. Hence, according to Eqs.\ref{eq:BS_q_dq} and \ref{eq:BS_l_dl}, we can set $\delta_l=\delta_q=r_0=2.4690386$ \r{A} for BCC Fe, and $\delta_l=\delta_q=2.4890153$ \r{A} for FCC Ni.  

Once $r_0$ is determined, we calculate $\alpha_q$ and $\gamma_q$ using Eqs. \ref{eq:E_mca_cub_K_q}, \ref{eq:E_mca_cub_K_dqdr} and \ref{eq:BS_q_dq}. Here, we set the experimental values of $K_1$ and $(1/K_1)(\partial K_1/\partial P)$ approximately at zero-temperature, that is, $K_1=55$ KJ/m$^3$ and $(1/K_1)(\partial K_1/\partial P)=-7.3\times10^{-2}$ GPa$^{-1}$ for BCC Fe, and $K_1=-126$ KJ/m$^3$ and $(1/K_1)(\partial K_1/\partial P)=-2.8\times10^{-2}$ GPa$^{-1}$ for FCC Ni \cite{Getz,SAWAOKA1975267}. As we see in Eq.\ref{eq:BS_q_dq}, we also need the bulk modulus. In principle we could set its experimental value or the one given by the EOS of this SD-MD model that was obtained in Section \ref{section:force_field}. In this work we choose the second option in order to describe more accurately the relation between volume and pressure of the SD-MD model. Inserting all these quantities in Eq.\ref{eq:BS_q_dq} via Eqs. \ref{eq:E_mca_cub_K_q} and \ref{eq:E_mca_cub_K_dqdr} leads to $\alpha_q=28.5189\mu$eV/atom and $\gamma_q=1.05331$ for BCC Fe, and $\alpha_q=-49.1335\mu$eV/atom and $\gamma_q=1.1186$ for FCC Ni.

Lastly, we calculate the Bethe-Slater parameters for the dipole term ($\alpha_l$ and $\gamma_l$) using Eqs. \ref{eq:l_dl} and \ref{eq:BS_l_dl}. In this case we need the values of the anisotropic magnetoelastic constants $b_1$ and $b_2$. These constants are related to the magnetostrictive coefficients ($\lambda_{001}$ and $\lambda_{111}$) and elastic constants ($c_{ij}$) via \cite{CLARK1980531,Cullen}
\begin{equation}
\begin{aligned}
b_1 & = -\frac{3}{2}\lambda_{001}(c_{11}-c_{12}),\\
b_2 & = -3\lambda_{111}c_{44}.
\label{eq:b1_b2}     
\end{aligned}
\end{equation}
To calculate $b_1$ and $b_2$ we use the experimental  magnetostrictive coefficients $\lambda_{001}=26\times10^{-6}$ and $\lambda_{111}=-30\times10^{-6}$ for BCC Fe, and  $\lambda_{001}=-60\times10^{-6}$ and $\lambda_{111}=-35\times10^{-6}$ for FCC Ni at zero-temperature \cite{Handley}. For the values of the elastic constants we choose the calculated ones with the SD-MD model including only the MEAM potential (see Table \ref{table:data_properties}). Doing so, we get $b_1=-3.74166$ MJ/m$^3$ and $b_2=10.4643$ MJ/m$^3$ for BCC Fe, and $b_1=10.0611$ MJ/m$^3$ and $b_2=13.9398$ MJ/m$^3$ for FCC Ni. If we insert these values in Eq.\ref{eq:BS_l_dl} via Eq.\ref{eq:l_dl}, then we obtain $\alpha_l=392.747\mu$eV/atom and $\gamma_l=0.824409$ for BCC Fe, and $\alpha_l=179.396\mu$eV/atom and $\gamma_l=1.39848$ for FCC Ni.

\begin{table}[h]
\caption{Parameters of the SD-MD model for BCC Fe and FCC Ni.}
\label{table:data_BS}
\centering
\begin{tabular}{ccc}
\toprule
\begin{tabular}[c]{@{}c@{}}\textbf{SD-MD model} \\ \textbf{parameters}\end{tabular}		& \quad\textbf{BCC Fe} \quad\quad &  \textbf{FCC  Ni} \\
\midrule
$\alpha_\mu$	($\mu_B^2\cdot$atom/\r{A}$^3$)					& 1.49057 & 0.172931 \\
$\beta_\mu$  ($\mu_B^2\cdot$atom$^2$/\r{A}$^6$)                    & -0.0978406 & -0.021997 \\
$\gamma_\mu$	($\mu_B^2\cdot$atom$^3$/\r{A}$^9$) 				& 0.0026366 & 0.00096755 \\
$v_{c}$ (\r{A}$^3$/atom)					& 6.39848 & 5.36535 \\
\midrule
$\alpha_J^{(I)}$ (meV/atom)					& -12.5921 & 8.35847  \\
$\gamma_J^{(I)}$                      & 2.81897  & -0.098217  \\
$\delta_J^{(I)}$ (\r{A})					&  2.4690386 & 2.4890153  \\
$R_{c,J}^{(I)}$ (\r{A})					& 2.6 & 2.6 \\
\midrule
$\alpha_J^{(II)}$ (meV/atom)					& 50.996$^a$ & 19.46$^a$ \\
$\gamma_J^{(II)}$                      & 0.281$^a$ & 0.00011$^a$  \\
$\delta_J^{(II)}$ (\r{A})					& 1.999$^a$ & 1.233$^a$ \\
$R_{c,J}^{(II)}$ (\r{A})					& 4.5$^a$ & 4.5$^a$ \\
\midrule
$\alpha_l$ ($\mu$eV/atom)					& 392.747 & 179.396  \\
$\gamma_l$                      & 0.824409 & 1.39848 \\
$\delta_l$ (\r{A})					& 2.4690386 & 2.4890153  \\
$R_{c,l}$ (\r{A})					& 2.6 & 2.6 \\
\midrule
$\alpha_q$	($\mu$eV/atom)					& 28.5189 & -49.1335 \\
$\gamma_q$                      & 1.05331 & 1.1186 \\
$\delta_q$	(\r{A})				& 2.4690386 & 2.4890153 \\
$R_{c,q}$ (\r{A})					& 2.6 & 2.6 \\
\bottomrule
$^a$Ref.\cite{TRANCHIDA2018406}
\end{tabular}
\end{table}

\begin{figure}[h]
\centering
\includegraphics[width=\columnwidth ,angle=0]{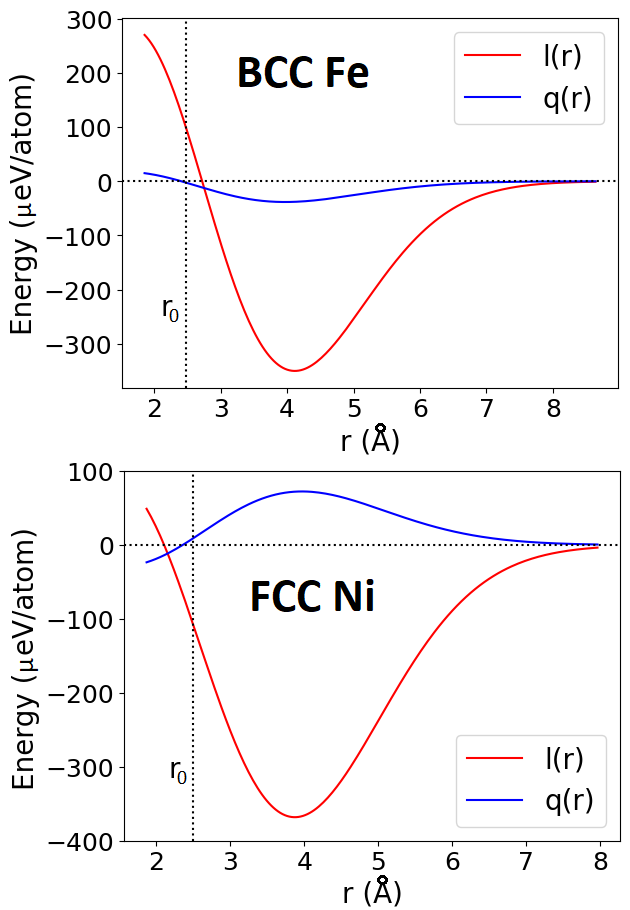}
\caption{Calculation of the Bethe-Slater function $l(r)$ and $q(r)$ for (top) BCC Fe and (bottom) FCC Ni using the parameters given in Table \ref{table:data_BS}. Vertical dash line stands for the equilibrium distance of the first nearest neighbors $r_0$.}
\label{fig:BS_Fe_Ni}
\end{figure}

The Bethe-Slater parameters for the constructed SD-MD models are shown in Table \ref{table:data_BS}, while the corresponding Bethe-Slater functions for $l(r)$ and $q(r)$ using these parameters are plotted in Fig.\ref{fig:BS_Fe_Ni}. We see that $l(r_0)$ is approximately two order of magnitude greater than $q(r_0)$. However, note that after taking into account all first nearest neighbors the N\'{e}el quadrupole and dipole energies can be of the same order of magnitude close to the cubic symmetry (see Section \ref{section:Neel_tests}). Fig.\ref{fig:BS_Fe_Ni} also contains interesting information about the dependence of MCA and magnetoelasticity on the distance between first nearest neighbors. For instance, we see that if we decrease the distance between first nearest neighbors from the equilibrium value $r_0$ (high hydrostatic pressure regime)  for both BCC Fe and FCC Ni, then the sign of $q(r)$ changes, which implies a change in the sign of $K_1$, see Eq.\ref{eq:E_mca_cub_K_q}.  Similarly, if we increase the distance between first nearest neighbors for BCC Fe ($r_0>3$\r{A}), then the sign of $l(r)$ changes switching the sign of $b_1$, see Eq.\ref{eq:l_dl}. In general, the physical interpretation of $q(r)$ and $l(r)$ far from the equilibrium value $r_0$ should be done with caution since we only involved up to the first derivative of these functions evaluated at $r_0$ in their parameterization. In this sense, the only meaningful region around $r_0$ may be where first order Taylor expansion at $r_0$ of the Bethe-Slater functions of $q(r)$ and $l(r)$ is a good approximation.  Including up the first derivative of $q(r)$ and $l(r)$ in their parameterization might be enough for many practical purposes since the distance between first nearest neighbors oscillates close to the equilibrium value at finite temperature below the melting point.

\begin{table*}[ht]
\caption{Calculated and experimental elastic constants, magnetostrictive coefficients, MCA, and MCA under hydrostatic pressure for BCC Fe and FCC Ni at zero-temperature.}
\label{table:data_properties}
\centering
\begin{tabular}{cccc|ccc|ccc|ccc}
\toprule
\textbf{Material} & \begin{tabular}[c]{@{}c@{}}\textbf{Elastic} \\ \textbf{constants}\end{tabular}	& \begin{tabular}[c]{@{}c@{}}\textbf{SD-MD} \\ \textbf{(GPa)}\end{tabular}	& \begin{tabular}[c]{@{}c@{}}\textbf{Expt.} \\ \textbf{(GPa)}\end{tabular}  & \begin{tabular}[c]{@{}c@{}}\textbf{Magnetostrictive} \\ \textbf{coefficients}\end{tabular}	& \begin{tabular}[c]{@{}c@{}}\textbf{SD-MD} \\ \textbf{($\times10^{-6}$)}\end{tabular}	& \begin{tabular}[c]{@{}c@{}}\textbf{Expt.} \\ \textbf{($\times10^{-6}$)}\end{tabular}  & \textbf{MCA} 	& \begin{tabular}[c]{@{}c@{}}\textbf{SD-MD} \\ \textbf{(KJ/m$^3$)}\end{tabular}	& \begin{tabular}[c]{@{}c@{}}\textbf{Expt.} \\ \textbf{(KJ/m$^3$)}\end{tabular} 
  & \textbf{MCA vs P} & \begin{tabular}[c]{@{}c@{}}\textbf{SD-MD} \\ \textbf{(GPa$^{-1}$)}\end{tabular}	& \begin{tabular}[c]{@{}c@{}}\textbf{Expt.} \\ \textbf{(GPa$^{-1}$)}\end{tabular}  \\

\midrule
BCC Fe					& $c_{11}$ & 230.0 &  230$^a$  & $\lambda_{001}$ & 25.9 & 26$^c$ &   $K_1$ & 54.995  &  55$^d$ &   $\frac{1}{K_1}\frac{\partial K_1}{\partial P}$ & -0.0727  &  -0.073$^e$\\
                      & $c_{12}$ & 134.1 & 135$^a$ & $\lambda_{111}$ & -30.3 & -30$^c$ &   &   &  &   &   & \\
					& $c_{44}$ & 116.3 & 117$^a$ & & & &  & & &   &   & \\
\midrule
FCC Ni					& $c_{11}$ & 263.9 &  261.2$^b$  & $\lambda_{001}$ & -61.9 & -60$^c$ &  $K_1$ & -125.996  &  -126$^d$ &  $\frac{1}{K_1}\frac{\partial K_1}{\partial P}$ & -0.0279 &  -0.028$^e$ \\
                      & $c_{12}$ & 152.1 & 150.8$^b$  & $\lambda_{111}$ & -35.4
                    & -35$^c$ &   &   &   &   &   & \\
					& $c_{44}$ & 132.8 & 131.7$^b$  & & & &  & & &   &   & \\
\bottomrule
$^a$Ref.\cite{asadi2015}, & $^b$Ref.\cite{Lee2003}, & $^c$Ref.\cite{Handley}, \\ $^d$Ref.\cite{Getz}, & $^e$Ref.\cite{SAWAOKA1975267}
\end{tabular}
\end{table*}

\section{Results}
\label{section:results}

\subsection{Tests of the N\'{e}el interaction}
\label{section:Neel_tests}

\begin{figure}[h]
\centering
\includegraphics[width=\columnwidth ,angle=0]{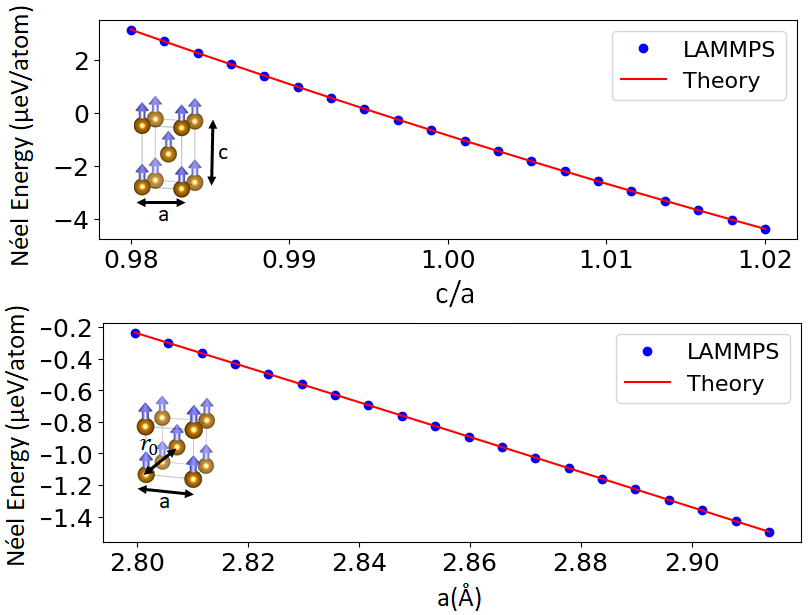}
\caption{Calculation of the N\'{e}el energy with LAMMPS and (top) Eq.\ref{eq:Neel_energy_dip} and (bottom) Eq.\ref{eq:Neel_energy_quad} for different values of the lattice parameters.}
\label{fig:neel_test}
\end{figure}

Before evaluating the magnetoelastic properties of the SD-MD model, it is convenient to check that the implementation of the N\'{e}el interaction Eq.\ref{eq:Neel_energy} in the SD-MD simulation is correct. To this end, we propose some tests by comparing the numerical results of the SD-MD simulation with simple analytical solutions. For instance, if we consider a BCC structure with N\'{e}el interactions up to first nearest neighbor in a collinear state along $\boldsymbol{s}=(0,0,1)$, then from Eq.\ref{eq:Neel_energy_coll} we have 
\begin{equation}
\begin{aligned}
\mathcal{H}_{N\acute{e}el} (0,0,1) & =   \frac{16Nq(r_0)}{45},
\label{eq:Neel_energy_quad}
\end{aligned}
\end{equation}
where $N$ is the number of atoms in the system, $r_0$ is the distance to nearest neighbor that is related to the lattice parameter $a$ via $r_0=a\sqrt{3}/2$. This equation allows to verify the quadrupole term. Let's now apply to this system with $\boldsymbol{s}=(0,0,1)$ a tetragonal deformation along the z-axis, where the lattice parameter is $c$ in this direction, and $a$ along both x-axis and y-axis. From Eq.\ref{eq:Neel_energy_coll} we obtain
\begin{equation}
\begin{aligned}
\mathcal{H}_{N\acute{e}el} (0,0,1) & =   -4Nl(r_0)\left[\frac{\left(\frac{c}{a}\right)^2}{2+\left(\frac{c}{a}\right)^2}-\frac{1}{3}\right]\\
& -\frac{16Nq(r_0)\left[2\left(\frac{c}{a}\right)^4-12\left(\frac{c}{a}\right)^2+3\right]}{35\left[2+\left(\frac{c}{a}\right)^2\right]^2},
\label{eq:Neel_energy_dip}
\end{aligned}
\end{equation}
where
\begin{equation}
\begin{aligned}
r_0 = \frac{a}{2}\sqrt{2+\left(\frac{c}{a}\right)^2}.
\label{eq:r0}
\end{aligned}
\end{equation}
This equation allows to check both the dipole and quadrupole terms. In the limit $c/a\xrightarrow{} 1$, the Eq.\ref{eq:Neel_energy_dip} becomes Eq.\ref{eq:Neel_energy_quad} ensuring the continuity of the N\'{e}el energy under structure deformation. In Fig.\ref{fig:neel_test}, we verify that the calculation of the N\'{e}el energy with LAMMPS is the same to   Eqs.\ref{eq:Neel_energy_quad} and \ref{eq:Neel_energy_dip} using the Bethe-Slater parameters of BCC Fe given in Table \ref{table:data_BS}. Similar tests could also be performed for other magnetic moment directions and deformations.

\subsection{Magnetic properties at zero-temperature}
\label{section:properties_tests}

In this section, we evaluate the magnetization and MCA under  pressure, anisotropic magnetostrictive coefficients, volume magnetostriction and saturation magnetization at zero-temperature given by the developed SD-MD models for BCC Fe and FCC Ni in Section \ref{section:SL_Fe_Ni}. We include MEAM potentials, exchange and N\'{e}el energies, and volume-dependent magnetic moment in the following calculations.  Magnetic collinear states will be used since we are interested in properties at zero-temperature.  All simulations are performed with the SPIN package of LAMMPS \cite{TRANCHIDA2018406}.

\subsubsection{Ground state}
\label{section:ground_state}

Firstly, we determine the equilibrium volume of the full SD-MD model (including the N\'{e}el interaction) for the conventional unit cell of BCC Fe and FCC Ni. To this end, we calculate the energy versus volume curve, and we fit it to the Murnaghan EOS in the same way as it was done in Fig.\ref{fig:eos} previously. Here, we also set the magnetic moments along the easy direction ($[1,0,0]$ for BCC Fe and $[1,1,1]$ for FCC Ni) in order to get the minimum energy of the quadrupole term of N\'{e}el interaction. The equilibrium volume found with this procedure is $v_0=11.5867635$ \r{A}$^3$/atom for BCC Fe, and $v_0=10.9035445$ \r{A}$^3$/atom for FCC Ni. We verify that pressure is lower than $5\times10^{-5}$ GPa in these equilibrium states. As we anticipated in Section \ref{section:SL_Fe_Ni}, the dipole and quadrupole N\'{e}el interactions induce a very small change in the equilibrium volume when is included in the SD-MD model.

\subsubsection{Magnetocrystalline anisotropy}

\begin{figure}[h]
\centering
\includegraphics[width=\columnwidth ,angle=0]{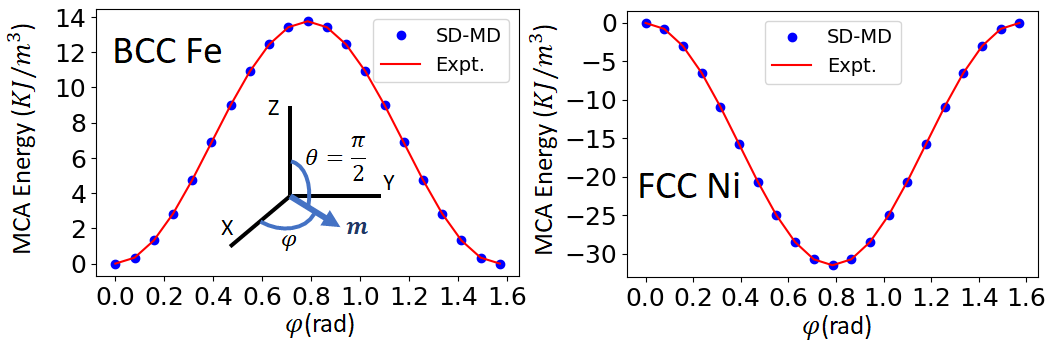}
\caption{Calculation of the MCA energy for BCC Fe and FCC Ni  with SD-MD simulation (blue points) and  Eq.\ref{eq:E_mca_cub} using the experimental $K_{1}$ (red line). Magnetic moments are constrained on the XY plane.}
\label{fig:mae_Fe_Ni}
\end{figure}

 Next, we compute the MCA energy at this equilibrium volume by setting the magnetic moment along different directions in the XY plane. In Fig.\ref{fig:mae_Fe_Ni} we show a comparison between  the  MCA energy  calculated by  SD-MD simulations with LAMMPS and  Eq.\ref{eq:E_mca_cub} using the experimental value ($K_{1}=55$ KJ/m$^3$ for BCC Fe and $K_{1}=-126$ KJ/m$^3$)\cite{Getz}.  The direct evaluation of $K_1$ with the SD-MD model through Eq.\ref{eq:E_mca_cub_K} gives $54.995$ KJ/m$^3$ for BCC Fe and  $-125.996$ KJ/m$^3$ for FCC Ni. As we see, the SD-MD model with the Bethe-Slater parameters given by Table \ref{table:data_BS} reproduces very well the first-order experimental MCA.

\begin{figure}[h]
\centering
\includegraphics[width=\columnwidth ,angle=0]{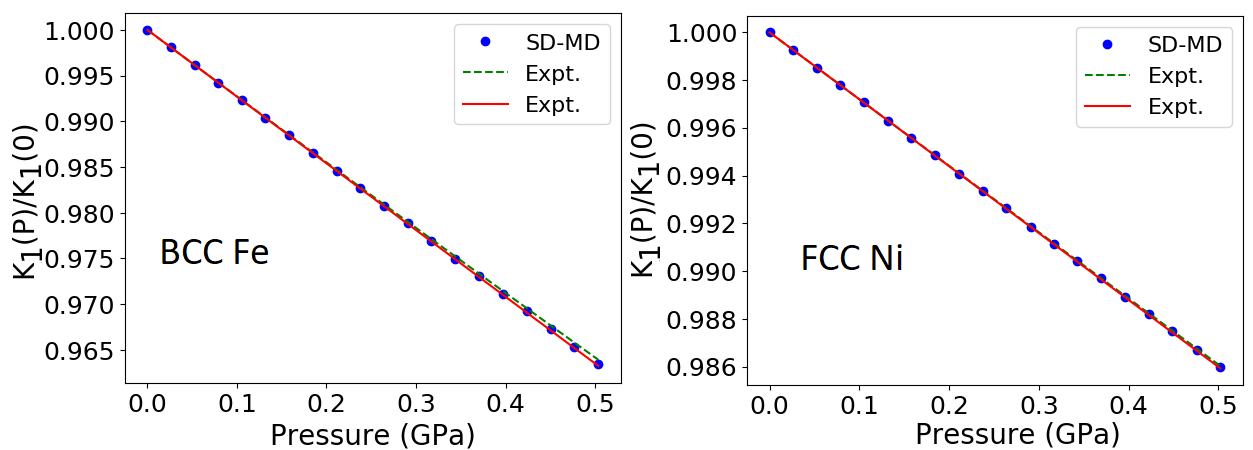}
\caption{Calculation of $K_1(P)/K_1(0)$ under hydrostatic pressure using the developed SD-MD model (blue dots) for BCC Fe and FCC Ni. The green and red lines stand for the experimental behaviour given by Eq.\ref{eq:dk_int_sol} and its low-pressure approximation Eq.\ref{eq:dk_int_sol_approx}, respectively \cite{SAWAOKA1975267}.}
\label{fig:mae_vs_press_Fe_Ni}
\end{figure}

Now we study the effects of hydrostatic pressure on the MCA for this SD-MD model. To facilitate the comparison between the model and experiment, we first convert $(1/K_1)(\partial K_1/\partial P)$ to an integral form, that is,
\begin{equation}
\begin{aligned}
\frac{1}{K_1}\frac{\partial K_1}{\partial P}=\zeta \quad\longrightarrow\quad \int_{K_1(0)}^{K_1(P)}\frac{dK_1}{K_1}=\int_0^P \zeta dP,
\label{eq:dk_int}
\end{aligned}
\end{equation}
where $\zeta=-7.3\times 10^{-2}$ GPa$^{-1}$ is the experimental value measured up to $P=0.5$ GPa at T=77K for BCC Fe\cite{SAWAOKA1975267}, while for FCC Ni is $\zeta=-2.8\times 10^{-2}$ GPa$^{-1}$. Solving this integral we have
\begin{equation}
\begin{aligned}
\frac{K_1(P)}{K_1(0)}=e^{\zeta P},
\label{eq:dk_int_sol}
\end{aligned}
\end{equation}
where in the low pressure regime ($\zeta P\ll 1$) it can be written as
\begin{equation}
\begin{aligned}
\frac{K_1(P)}{K_1(0)}\approx 1+\zeta P+O(P^2).
\label{eq:dk_int_sol_approx}
\end{aligned}
\end{equation}
In Fig.\ref{fig:mae_vs_press_Fe_Ni} we show the ratio $K_1(P)/K(0)$ versus pressure generated by the SD-MD model of Fe and Ni, and the experimental behaviour given by Eq.\ref{eq:dk_int_sol} and its low-pressure approximation Eq.\ref{eq:dk_int_sol_approx}. The linear fitting to the data generated by the SD-MD model up to $P=0.5$ GPa gives $(1/K_1)(\partial K_1/\partial P)=-7.27\times 10^{-2}$ GPa$^{-1}$ for BCC Fe, and $-2.79\times 10^{-2}$ GPa$^{-1}$ for FCC Ni, which is in very good agreement with the experimental values\cite{SAWAOKA1975267}. Note that Eq.\ref{eq:dk_int_sol} and MCA results of the model beyond the range of pressure between $0$GPa and $0.5$GPa should be taken with caution due to the lack of experimental data.

\begin{figure}[h]
\centering
\includegraphics[width=\columnwidth ,angle=0]{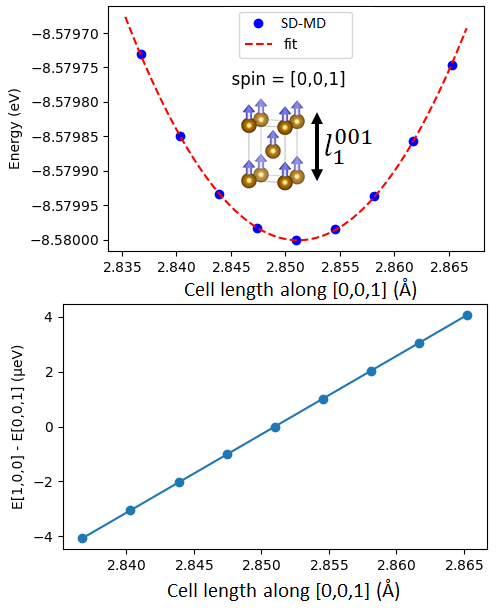}
\caption{Calculation of $\lambda_{001}$ for BCC Fe using MAELAS interfaced with LAMMPS. (top) Quadratic curve fit to the energy versus cell length along $\boldsymbol{\beta}=\left(0,0,1\right)$ with spin direction $\boldsymbol{s}_1=\left(0,0,1\right)$ under a volume-conserving tetragonal deformation. (bottom) Energy difference between states with spin directions $\boldsymbol{s}_2=\left(1,0,0\right)$ and $\boldsymbol{s}_1=\left(0,0,1\right)$ against the cell length along $\boldsymbol{\beta}=\left(0,0,1\right)$.}
\label{fig:lmb_001_Fe}
\end{figure}

\begin{figure}[h]
\centering
\includegraphics[width=\columnwidth ,angle=0]{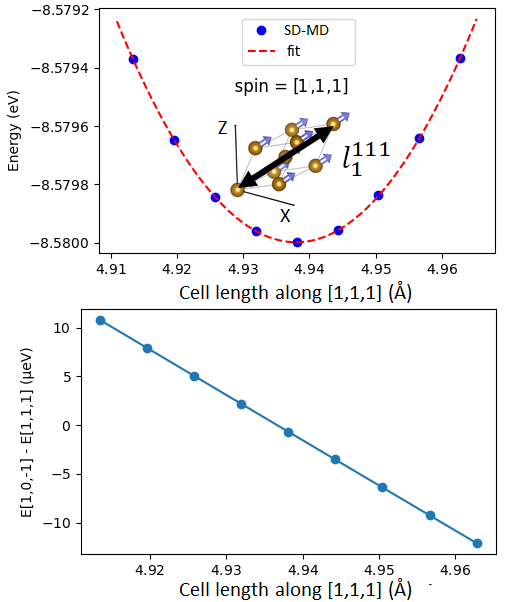}
\caption{Calculation of $\lambda_{111}$ for BCC Fe using MAELAS interfaced with LAMMPS. (top) Quadratic curve fit to the energy versus cell length along $\boldsymbol{\beta}=\left(1/\sqrt{3},1/\sqrt{3},1/\sqrt{3}\right)$ with spin direction $\boldsymbol{s}_1=\left(1/\sqrt{3},1/\sqrt{3},1/\sqrt{3}\right)$ under a volume-conserving trigonal deformation. (bottom) Energy difference between states with spin directions $\boldsymbol{s}_2=\left(1/\sqrt{2},0,-1/\sqrt{2}\right)$ and $\boldsymbol{s}_1=\left(1/\sqrt{3},1/\sqrt{3},1/\sqrt{3}\right)$ against the cell length along $\boldsymbol{\beta}=\left(1/\sqrt{3},1/\sqrt{3},1/\sqrt{3}\right)$.}
\label{fig:lmb_111_Fe}
\end{figure}

\subsubsection{Anisotropic magnetostriction}

Now, we compute the anisotropic magnetostrictive coefficients using the SD-MD model. To this end, we apply the method proposed by Wu and Freeman \cite{Wu1996,WU1997} as implemented in the program MAELAS\cite{nieves2020maelas,Maelas}.  In this method, the anisotropic magnetostrictive coefficients for cubic systems (point groups $432$, $\bar{4}3m$, $m\bar{3}m$) are calculated as \cite{nieves2020maelas}
\begin{equation}
     \lambda_{001}=\frac{4 (l^{001}_1 -l^{001}_2)}{3(l^{001}_1 +l^{001}_2)},\quad \lambda_{111}=\frac{4 (l^{111}_1 -l^{111}_2)}{3(l^{111}_1 +l^{111}_2)},
    \label{eq:lambda_method}
\end{equation}
where $l^{001}_1$ and $l^{001}_2$ are the equilibrium cell lengths along the length measuring direction $\boldsymbol{\beta}=(0,0,1)$ under a tetragonal deformation with collinear magnetic moment directions $\boldsymbol{s}_1=\left(0,0,1\right)$ and $\boldsymbol{s}_2=\left(1,0,0\right)$, respectively. Similarly, $l^{111}_1$ and $l^{111}_2$ are the equilibrium cell lengths along the length measuring direction $\boldsymbol{\beta}=(1/\sqrt{3},1/\sqrt{3},1/\sqrt{3})$ under a trigonal deformation with magnetic moment direction $\boldsymbol{s}_1=\left(1/\sqrt{3},1/\sqrt{3},1/\sqrt{3}\right)$ and $\boldsymbol{s}_2=\left(1/\sqrt{2},0,-1/\sqrt{2}\right)$, respectively. In order to obtain the equilibrium cell lengths $l^{001}_1$ and $l^{001}_2$, one needs to evaluate the energy for a set of volume-conserving tetragonal distorted unit cells. Next, the energy versus the cell length along $\boldsymbol{\beta}=(0,0,1)$ for each magnetic moment direction $\boldsymbol{s}_1=\left(0,0,1\right)$ and $\boldsymbol{s}_2=\left(1,0,0\right)$ is fitted to a quadratic function
\begin{equation}
      E(x)\Bigg\vert_{\boldsymbol{\beta}=(0,0,1)}^{\boldsymbol{s}_j}=\Tilde{a}_j x^2+\Tilde{b}_j x+\Tilde{c}_j,\quad j=1,2 
    \label{eq:E_fit}
\end{equation}
where $\Tilde{a}_j$, $\Tilde{b}_j$ and $\Tilde{c}_j$ are fitting parameters. The minimum of this quadratic function for magnetic moment direction $\boldsymbol{s}_{1(2)}$ corresponds to  $l^{001}_{1(2)}=-\Tilde{b}_{1(2)}/(2\Tilde{a}_{1(2)})$, and it is the equilibrium cell length. Similarly, the equilibrium cell lengths $l^{111}_1$ and $l^{111}_2$ are obtained by applying a set of volume-conserving trigonal deformations, and performing a quadratic fitting of the energy versus the cell length along $\boldsymbol{\beta}=\left(1/\sqrt{3},1/\sqrt{3},1/\sqrt{3}\right)$ with magnetic moment directions $\boldsymbol{s}_1=\left(1/\sqrt{3},1/\sqrt{3},1/\sqrt{3}\right)$ and $\boldsymbol{s}_2=\left(1/\sqrt{2},0,-1/\sqrt{2}\right)$.

\begin{figure}[h]
\centering
\includegraphics[width=\columnwidth ,angle=0]{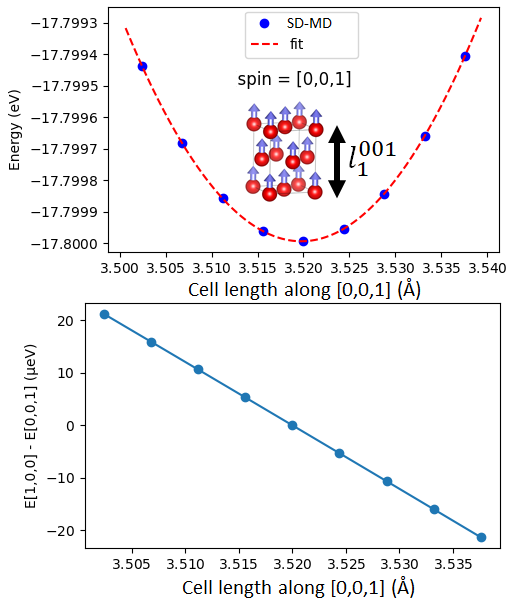}
\caption{Calculation of $\lambda_{001}$ for FCC Ni using MAELAS interfaced with LAMMPS. (top) Quadratic curve fit to the energy versus cell length along $\boldsymbol{\beta}=\left(0,0,1\right)$ with spin direction $\boldsymbol{s}_1=\left(0,0,1\right)$ under a volume-conserving tetragonal deformation. (bottom) Energy difference between states with spin directions $\boldsymbol{s}_2=\left(1,0,0\right)$ and $\boldsymbol{s}_1=\left(0,0,1\right)$ against the cell length along $\boldsymbol{\beta}=\left(0,0,1\right)$.}
\label{fig:lmb_001_Ni}
\end{figure}

\begin{figure}[h]
\centering
\includegraphics[width=\columnwidth ,angle=0]{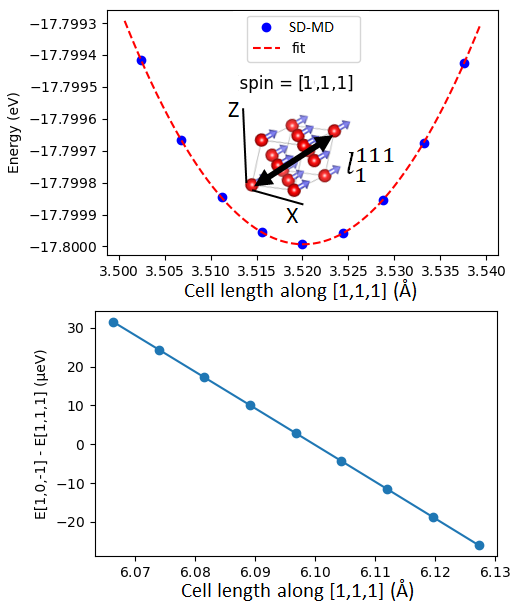}
\caption{Calculation of $\lambda_{111}$ for FCC Ni using MAELAS interfaced with LAMMPS. (top) Quadratic curve fit to the energy versus cell length along $\boldsymbol{\beta}=\left(1/\sqrt{3},1/\sqrt{3},1/\sqrt{3}\right)$ with spin direction $\boldsymbol{s}_1=\left(1/\sqrt{3},1/\sqrt{3},1/\sqrt{3}\right)$ under a volume-conserving trigonal deformation. (bottom) Energy difference between states with spin directions $\boldsymbol{s}_2=\left(1/\sqrt{2},0,-1/\sqrt{2}\right)$ and $\boldsymbol{s}_1=\left(1/\sqrt{3},1/\sqrt{3},1/\sqrt{3}\right)$ against the cell length along $\boldsymbol{\beta}=\left(1/\sqrt{3},1/\sqrt{3},1/\sqrt{3}\right)$.}
\label{fig:lmb_111_Ni}
\end{figure}

We have developed an interface between the software MAELAS\cite{nieves2020maelas} and LAMMPS\cite{TRANCHIDA2018406} in order to apply this method and extract the magnetostrictive coefficients easily. This interface is publicly available on GitHub repository \cite{maelas_lmp}. In Fig.\ref{fig:lmb_001_Fe} we show the quadratic curve fit to the energy versus cell length along $[0,0,1]$ with magnetic moment direction $\boldsymbol{s}_1=\left(0,0,1\right)$ to calculate $\lambda_{001}$ for BCC Fe. We also plot the energy difference between states with spin directions $\boldsymbol{s}_1=\left(1,0,0\right)$ and $\boldsymbol{s}_2=\left(0,0,1\right)$ against the cell length along $[0,0,1]$.  The corresponding plot for $\lambda_{111}$ is presented in Fig.\ref{fig:lmb_111_Fe}. We obtain $\lambda_{001}=25.9\times10^{-6}$ and $\lambda_{111}=-30.3\times10^{-6}$, while the experimental values\cite{Handley} at $T=4.2$K are $\lambda_{001}=26\times10^{-6}$ and $\lambda_{111}=-30\times10^{-6}$. The results for FCC Ni are plotted in Figs. \ref{fig:lmb_001_Ni} and \ref{fig:lmb_111_Ni}. Here, we get $\lambda_{001}=-61.9\times10^{-6}$ and $\lambda_{111}=-35.4\times10^{-6}$, while the experimental values\cite{Handley} at $T=4.2$K are $\lambda_{001}=-60\times10^{-6}$ and $\lambda_{111}=-35\times10^{-6}$. Therefore, the developed SD-MD model for Fe and Ni also exhibits magnetostrictive properties very similar to the experiment. Additionally, this calculation reveals that  the method proposed by Wu and Freeman \cite{Wu1996,WU1997} is an excellent approach to obtain the magnetostrictive coefficients as long as both the elastic and magnetoelastic energies are properly described by the model. This fact could not be verified before for $\lambda_{111}$ of BCC Fe due to a possible failure of  Density Function Theory calculations \cite{guo2002,Jones2015,Burkert,nieves2020maelas}. In Table \ref{table:data_properties} we present a summary of the results given by the SD-MD model for the MCA, MCA under hydrostatic pressure, and anisotropic magnetostrictive coefficients.

\begin{table}[h!]
\caption{Calculated volume magnetostriction $\omega_s$ with the SD-MD model for BCC Fe and FCC Ni using the set I and II of parameters in Table \ref{table:data_BS} to describe $J(r)$. Theoretical and experimental results found in literature are also shown for comparison.}
\label{table:data_ws}
\centering
\begin{tabular}{ccccc}
\toprule
\quad & \begin{tabular}[c]{@{}c@{}}\textbf{SD-MD} \\ \textbf{set I}\\ ($\times10^{-4}$)\end{tabular}	\quad	& \begin{tabular}[c]{@{}c@{}}\textbf{SD-MD} \\ \textbf{set II}\\ ($\times10^{-4}$)\end{tabular}	\quad &  \begin{tabular}[c]{@{}c@{}}\textbf{Theory} \\ \\ ($\times10^{-4}$)\end{tabular}	\quad & \begin{tabular}[c]{@{}c@{}}\textbf{Expt.} \\ \\ ($\times10^{-4}$)\end{tabular} \\
\midrule
BCC Fe				& 118 & -235 & 116$^a$ & 4$^b$ \\
                     &   &  & 683$^c$ & \\
\midrule
FCC Ni				& 3.71 & -53.7 & 3.75$^a$ & 3.65$^e$ \\
                     &   &  & 45.7$^c$ & 3.24$^f$ \\
                      &   &  &  & -5.1$^d$ \\
                            &   &  &  & -2.7$^b$ \\
\bottomrule
$^a$Ref.\cite{shimizu1978}, $^b$Ref.\cite{richter},\\ $^c$Ref.\cite{Janak1976},$^d$Ref.\cite{tanji},\\ $^e$Ref.\cite{nix}, 
$^f$Ref.\cite{williams}
\end{tabular}
\end{table}

\subsubsection{Volume magnetostriction}
\label{section:vol_mag}

\begin{figure}[h]
\centering
\includegraphics[width=\columnwidth ,angle=0]{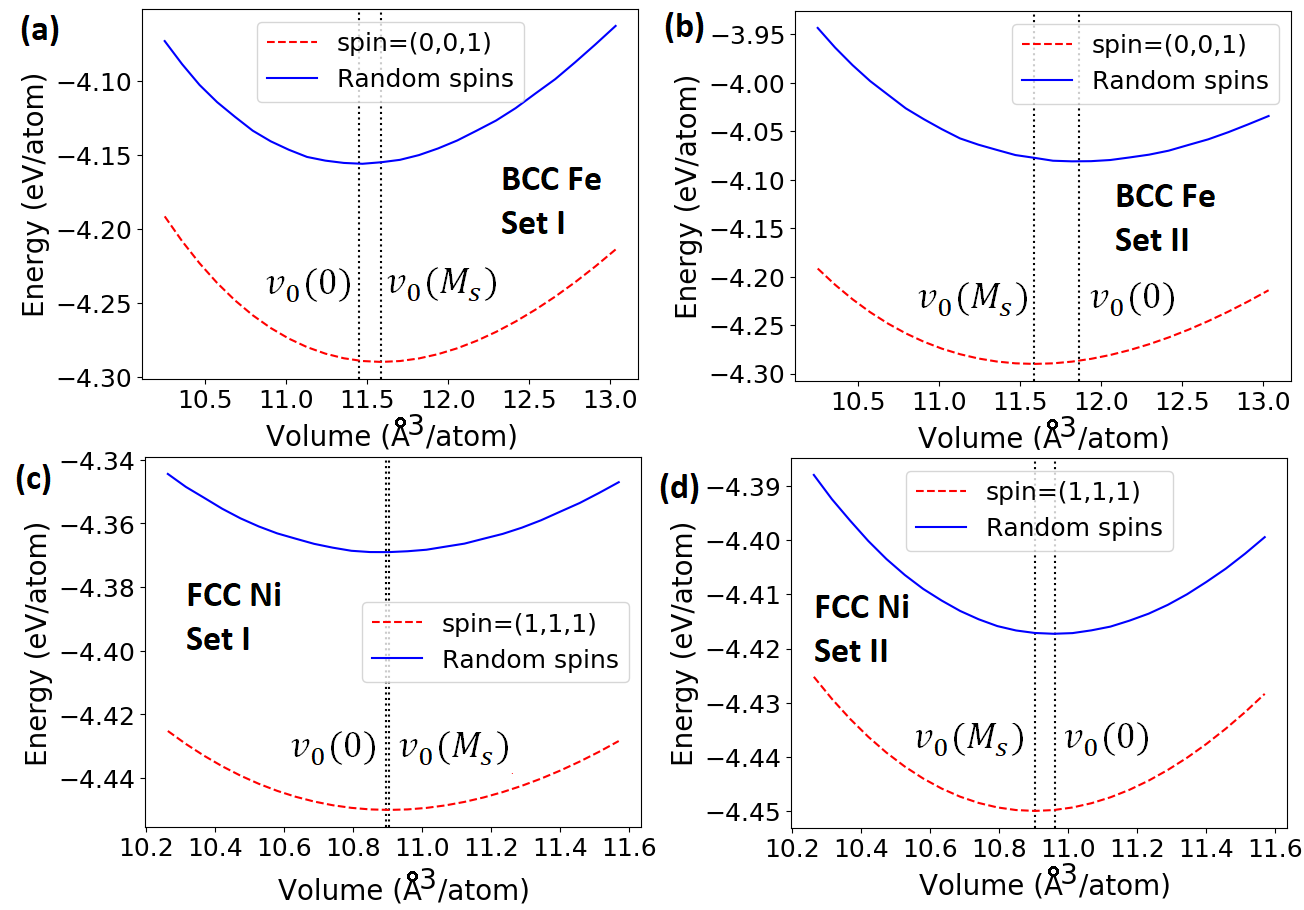}
\caption{Calculation of $\omega_s$ with the SD-MD model for ((a)-(b)) BCC Fe and ((c)-(d)) FCC Ni using the two set of parameters given in Table \ref{table:data_BS} to describe the exchange interaction $J(r)$. }
\label{fig:ws}
\end{figure}

The volume magnetostriction is generated by the presence of ferromagnetism in the magnetic material (exchange magnetostriction). It can be calculated as\cite{Khmele2004}
\begin{equation}
      \omega_s (T)= \frac{v_0(M_s(T))-v_0(0)}{v_0(0)},
    \label{eq:ws}
\end{equation}
where $v_0(M_s(T))$ and $v_0(0)$ are the equilibrium volume per atom in the magnetized and demagnetized (paramagnetic) states, respectively. In the magnetized state, the magnetization is equal to the saturation magnetization $M_s$ at temperature $T$. Hence, the quantity $v_0(M_s(T))$ at zero-temperature was already calculated in Section \ref{section:ground_state}. To compute $v_0(0)$ we apply a similar procedure. Namely, we first calculate the energy of a supercell with magnetic moments oriented randomly (demagnetized state) for different values of the lattice parameter $a$, preserving the cubic crystal symmetry. Next, we fit the energy versus volume curve to the Murnaghan EOS.  We use a supercell with 20x20x20 conventional unit cells with periodic boundary conditions for both BCC Fe (16000 atoms)  and FCC Ni (32000 atoms). We perform this calculation using the set I and II of parameters given in Table \ref{table:data_BS} to describe the exchange interaction $J(r)$. The results are depicted in Fig. \ref{fig:ws}. The set I gives $\omega_s=1.18\times10^{-2}$ for BCC Fe and $3.71\times10^{-4}$ for FCC Ni, reproducing fairly well the theoretical values calculated by Shimizu \cite{shimizu1978} ($\omega_s=1.16\times10^{-2}$ for BCC Fe and $3.75\times10^{-4}$ for FCC Ni) that we used to compute the Bethe-Slater parameters for $J(r)$ in Section \ref{section:exchange}. The set II leads to $\omega_s=-2.23\times10^{-2}$ for BCC Fe and $-5.37\times10^{-3}$ for FCC Ni, so they have the opposite sign to the results given by set I. According to Eq. \ref{eq:J_dJ}, these results may be understood in terms of $\partial J/\partial r$ at the first-nearest neighbors ($r=r_0$) since $\omega_s\propto\partial J/\partial r$. In Fig.\ref{fig:BS_J_Fe_Ni}, we observe that set I gives $\partial J/\partial r>0$ at $r=r_0$ for both Fe and Ni, while set II gives $\partial J/\partial r<0$ at $r=r_0$. Note that Eq. \ref{eq:J_dJ} is derived assuming only exchange interactions up to first-nearest neighbors, and set II has a large cut-off that includes exchange interactions beyond first-nearest neighbors. Wang et al. performed first-principles calculations of $J(r)$ finding a change in the sign of $\partial J/\partial r$ close to $r=r_0$, and $\partial J/\partial r>0$ for the second nearest neighbors \cite{Wang2010}. Previous theoretical and experimental works reported a positive volume magnetostriction for BCC Fe\cite{Janak1976,richter,shimizu1978}, while for FCC Ni one can find contradictory results with positive\cite{shimizu1978,Janak1976,nix,williams} and negative\cite{richter,tanji} values. A summary of these results is presented in Table \ref{table:data_ws}. As seen in Fig.\ref{fig:BS_J_Fe_Ni}, there is a maximum of $J(r)$ close to $r_0$ for Ni using the set I, so that a small increase in the lattice parameter would change the sign of $\partial J/\partial r$, and consequently the sign of $\omega_s$. Lastly, we point out that the isotropic magnetostrictive coefficient of cubic crystals ($\lambda^{\alpha}$) and magnetoelastic constant $b_0$ are related to the volume magnetostriction as\cite{ANDREEV199559,nieves2020maelas}
\begin{equation}
      \lambda^{\alpha}=\frac{-b_0-\frac{1}{3}b_1}{c_{11}+2c_{12}}=\frac{\omega_s}{3}.
    \label{eq:lamb_iso}
\end{equation}
Hence, we see that the isotropic magnetostriction is greater than the anisotropic one for both BCC Fe and FCC Ni.

\subsubsection{Saturation magnetization}

\begin{figure}[h]
\centering
\includegraphics[width=\columnwidth ,angle=0]{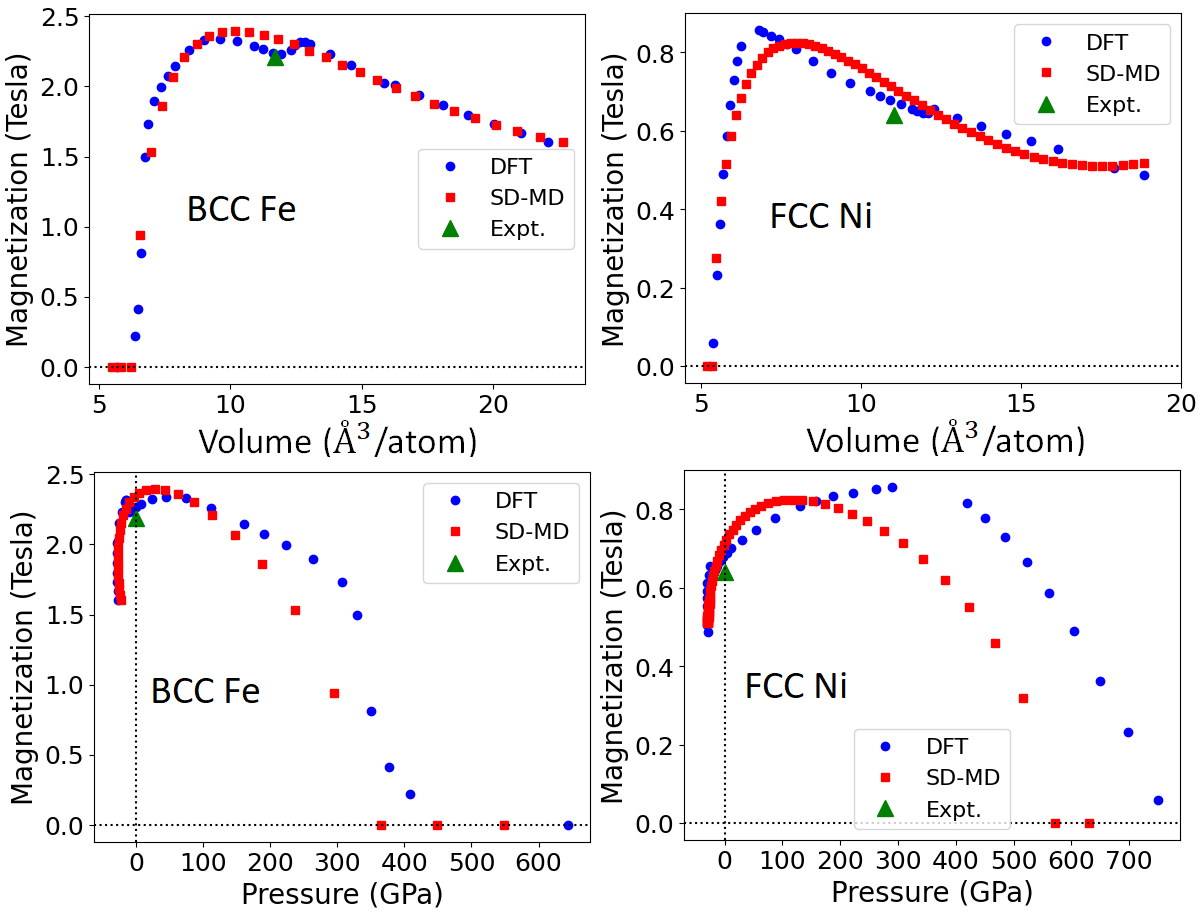}
\caption{Saturation magnetization against volume and pressure calculated with the SD-MD model including a volume-dependent magnetic moment (red squares) for BCC Fe and FCC Ni. Blue dots and green triangles stand for DFT and experimental data, respectively.}
\label{fig:mag_vs_press_Fe_Ni}
\end{figure}

The saturation magnetization at zero-temperature is computed using the following equation
\begin{equation}
\begin{aligned}
\mu_0M_s(v)=\frac{\mu_0\mu(v)}{v},
\label{eq:msat}
\end{aligned}
\end{equation}
where $\mu(v)$ is calculated using the Eq.\ref{eq:magmom_final} with the parameters shown in Table \ref{table:data_BS}. At the equilibrium volume of the SD-MD model it gives $\mu_0M_s(v_0)=2.35$T for BCC Fe, and $0.71$T for FCC Ni. The experimental values at zero-temperature are $\mu_0M_s=2.19$T for BCC Fe, and $0.64$T for FCC Ni\cite{Handley}. We see that the model slightly overestimates the saturation magnetization. Next, we  evaluate $M_s$ for different volumes applying normal deformations. The results of this calculation are shown in Fig.\ref{fig:mag_vs_press_Fe_Ni}. Here, we also included the data given by DFT that we obtained in Section \ref{section:mag_mom}. We observe that the overall behaviour of $M_s$ is well described by the model.  As we increase the volume above the equilibrium volume $v_0$, the pressure becomes negative and $M_s$ is decreasing. The condition that causes $M_s$ to decrease with volume is
\begin{equation}
\begin{aligned}
\frac{\partial M_s}{\partial v}<0\quad\longrightarrow \quad \frac{\partial\mu}{\partial v}<\frac{\mu}{v}.
\label{eq:msat_decrease}
\end{aligned}
\end{equation}
On the other hand, if we decrease the volume below the equilibrium volume then the pressure is positive. At high positive pressure, $M_s$ becomes zero when the volume per atom is lower than the critical volume ($v<v_c$) where magnetic moment collapses ($\mu=0$).

\section{Conclusions}
\label{section:con}

Many aspects of magnetoelastic phenomena are not fully understood yet due to the complexity of the materials at large scale.  Advanced modeling techniques and associated numerical tools based on a bottom-up multiscale approach could help to get a better understanding of magnetoelastic phenomena in magnetic materials across length scales. In this sense the SD-MD simulations using the N\'{e}el model could play an important role linking the atomic and macroscopic scales. Aiming at exploring this possibility, we showed a general methodology to build SD-MD models to describe  MCA under hydrostatic pressure, anisotropic magnetostriction, volume magnetostriction and saturation magnetization. To illustrate the method, we successfully applied it to BCC Fe and FCC Ni at zero-temperature. 

We aim at transposing our methodology to other materials and crystal structures. 
For example, in magnetic oxides and 4-f magnets, the MCA can correspond to energies orders of magnitude larger than in magnetic 3-d metals. 
Our approach could be used to map the subsequent interactions and build meso-scale models that will help reveal the influence of magnetism on large-scale thermo-elastic materials properties. Possible extensions of these models might be also useful to study morphic effects \cite{Rouchy1979,DUTREMOLETDELACHEISSERIE198277}. 

Although this work focused on bulk magnetoelasticity, similar effects have been shown to be important for smaller scale, finite-size systems \cite{Beaujouan2012}.
Previous studies have been investigating the relevance of the N\'{e}el model to simulate surface effects in 3-d magnetic metals \cite{yanes2007effective,skomski1998magnetoelectric}.
Future work could leverage our framework to develop surface interaction models for spin-lattice simulations of magnetic nanoparticles \cite{dos2020size}, as well as models for magnetic alloys \cite{Neel,Chika}.

The results presented in this work also raise interesting questions for future research on how these models will perform at finite temperature, and under magnetic field and stress. In particular, it would be interesting to study the possible correlations between the thermal variation of the magnetoelastic constants and the magnetization given by these models \cite{DUTREMOLETDELACHEISSERIE1983837,Evans,Evans2020}.

\section*{Acknowledgement}

This work was supported by the ERDF in the IT4Innovations national supercomputing center - path to exascale project (CZ.02.1.01/0.0/0.0/16-013/0001791) within the OPRDE.
This work was supported by The Ministry of Education, Youth and Sports from  the Large Infrastructures for Research, Experimental Development, and Innovations project “e-INFRA CZ - LM2018140”. This work was also supported by the Donau project No. 8X20050. P.N., D.L., and~S.A. acknowledge support from the H2020-FETOPEN no.~863155 s-NEBULA project. Sandia National Laboratories is a multimission laboratory managed and operated by National Technology and Engineering Solutions of Sandia, LLC., a wholly owned subsidiary of Honeywell International, Inc., for the U.S. Department of Energy's National Nuclear Security Administration under contract DE-NA-0003525. This paper describes objective technical results and analysis. Any subjective views or opinions that might be expressed in the paper do not necessarily represent the views of the U.S. Department of Energy or the United States Government.

\medskip

\section*{Appendix}

\appendix

\section{Derivation of $\partial q/\partial r$}
\label{App_dqdr}

In this appendix we show the steps to obtain the final expression  for $\partial q/\partial r$ given by Eq. \ref{eq:E_mca_cub_K_dqdr}. Firstly, we write the derivative of $q(r)$ with respect to the first nearest neighbor distance $r$ in Eq. \ref{eq:E_mca_cub_K_q} as
\begin{equation}
\begin{aligned}
 q(r) = -\frac{\xi VK_1}{N} \longrightarrow  r_0\frac{\partial q}{\partial r}\Big\vert_{r=r_0} = - \frac{\xi r_0}{N}\frac{\partial (VK_1)}{\partial r}\Big\vert_{r=r_0} ,
\label{eq:app_q}     
\end{aligned}
\end{equation}
where $r_0$ is the equilibrium distance to the first nearest neighbors, $N$ is the number of atoms in the volume $V$,  and $\xi$ is equal to $-1/2$, $9/16$, and $1$ for SC, BCC and FCC, respectively. Next, we work out this equation in the following way
\begin{equation}
\begin{aligned}
 r_0\frac{\partial q}{\partial r}\Big\vert_{r=r_0} & = - \frac{\xi r_0}{N}\frac{\partial (VK_1)}{\partial r}\Big\vert_{r=r_0} = - \frac{\xi r_0}{N}\left[K_1\frac{\partial V}{\partial r}+V\frac{\partial K_1}{\partial r}\right]_{r=r_0}\\ 
& = - \frac{\xi r_0}{N}\frac{\partial V}{\partial r}\Big\vert_{r=r_0}\left[K_1+V\frac{\partial K_1}{\partial V}\right]_{r=r_0}\\
& = - \frac{\xi r_0}{N}\frac{\partial V}{\partial r}\Big\vert_{r=r_0}\left[K_1+V\frac{\partial P}{\partial V}\frac{\partial K_1}{\partial P}\right]_{r=r_0},
\label{eq:app_dq}     
\end{aligned}
\end{equation}
where
\begin{equation}
\begin{aligned}
& SC: \quad\frac{\xi r_0}{N} \frac{\partial V}{\partial r} \Big\vert_{r=r_0}= -\frac{3}{2}r_0^3,\\
& BCC:\quad \frac{\xi r_0}{N} \frac{\partial V}{\partial r} \Big\vert_{r=r_0}= \frac{3\sqrt{3}}{4}r_0^3,\\
& FCC:\quad \frac{\xi r_0}{N} \frac{\partial V}{\partial r} \Big\vert_{r=r_0}= \frac{3}{\sqrt{2}}r_0^3.
\label{eq:app_dVdr}     
\end{aligned}
\end{equation}

Lastly, we make use of the definition of the bulk modulus $B=-V(\partial P/\partial V)$ in Eq.\ref{eq:app_dq}. Doing so, we obtain 

\begin{equation}
\begin{aligned}
& SC: r_0\frac{\partial q}{\partial r}\Big\vert_{r=r_0}  = \frac{3}{2}r_0^3K_1(r_0)\left[1-\frac{B}{K_1}\frac{\partial K_1}{\partial P}\right]_{r=r_0},\\
& BCC: r_0\frac{\partial q}{\partial r}\Big\vert_{r=r_0}  = - \frac{3\sqrt{3}}{4}r_0^3K_1(r_0)\left[1-\frac{B}{K_1}\frac{\partial K_1}{\partial P}\right]_{r=r_0},\\
& FCC: r_0\frac{\partial q}{\partial r}\Big\vert_{r=r_0}  = - \frac{3}{\sqrt{2}}r_0^3K_1(r_0)\left[1-\frac{B}{K_1}\frac{\partial K_1}{\partial P}\right]_{r=r_0}.
\label{eq:E_mca_cub_K_dqdr_appendix}     
\end{aligned}
\end{equation}

\bibliography{mybibfile.bib}
 

\end{document}